\documentclass[twocolumn,prl,superscriptaddress,floats,nobibnotes,notitlepage,citeautoscript]{revtex4-1}

\usepackage[pdftex]{graphicx,hyperref}
\usepackage[utf8]{inputenc}
\usepackage{amsmath,bm}
\usepackage{bbold}
\usepackage{color,soul}
\usepackage{xstring}
\usepackage{wasysym}
\usepackage{xspace}
\usepackage{amssymb}
\usepackage{time}
\usepackage{bbold}
\usepackage{subfigure}
\usepackage{multirow}
\usepackage{xspace}
\usepackage{url}
\usepackage{float}
\usepackage{booktabs}
\usepackage{array}
\UseRawInputEncoding

\def\<{\langle}
\def\>{\rangle}

\newcommand{\ve}[1]{\boldsymbol{#1}}

\makeatother

\makeindex

\begin{document}

\title{Effective  model   for  superconductivity in magic-angle graphene}  

\author{Disha Hou}
\affiliation{\mbox{Department of Physics, Beijing Normal University, Beijing 100875, China}}
\author{\firstname{Yuhai} \surname{Liu}}
\affiliation{\mbox{School of Science, Beijing University of Posts and Telecommunications, Beijing 100876, China}}
\author{ Toshihiro Sato}
\affiliation{\mbox{Institut f\"ur Theoretische Physik und Astrophysik, Universit\"at W\"urzburg, 97074 W\"urzburg, Germany}}
\author{Fakher F. Assaad}
\email{assaad@physik.uni-wuerzburg.de}
\affiliation{\mbox{Institut f\"ur Theoretische Physik und Astrophysik, Universit\"at W\"urzburg, 97074 W\"urzburg, Germany}}
\affiliation{\mbox{W\"urzburg-Dresden Cluster of Excellence ct.qmat, Am Hubland, 97074 W\"urzburg, Germany}}
\author{\firstname{Wenan} \surname{Guo}}
\email{waguo@bnu.edu.cn}
\affiliation{\mbox{Department of Physics, Beijing Normal University, Beijing 100875, China}}
\affiliation{\mbox{Beijing Computational Science Research Center, 10 East Xibeiwang Road, Beijing 100193, China}}
\author{Zhenjiu Wang }
\email{Zhenjiu.Wang@physik.uni-muenchen.de}
\affiliation{Arnold Sommerfeld Center for Theoretical Physics, University of Munich, Theresienstr. 37, 80333 Munich, Germany}
\affiliation{Max-Planck-Institut f\"ur Physik komplexer Systeme, Dresden 01187, Germany}

\begin{abstract}
We  carry  out large-scale quantum Monte Carlo simulations of a candidate field theory  for  the  onset of  superconductivity  in 
 magic-angle  twisted bilayer graphene.   The  correlated   insulating  state   at  charge neutrality 
 spontaneously breaks   U(1)  Moir\'e valley   symmetry.   Owing to the   topological nature  of  the bands,   skyrmion  defects 
of  the   order parameter  carry charge  $2e$  and  condense  upon  doping.  In our calculations we  encode the  U(1)  symmetry by 
 an  internal degree of freedom  such  that  it is not  broken upon lattice  regularization.   Furthermore,  the  skyrmion carries  the same  charge. 
The  nature of  the  doping-induced phase transitions depends on the strength of   the  easy-plane anisotropy  that  
reduces  the  SU(2)   valley   symmetry  to  U(1) $\times   \mathbb{Z}_2 $. 
For  large  anisotropy,  we  observe  two  distinct transitions separated by  phase coexistence.  
While  the  insulator  to  superconducting transition is of  mean-field character,  the U(1)  transition   
is consistent  with  three-dimensional  XY  criticality.   Hence,  the coupling between the gapless charge 
excitations of the superconducting phase and   the XY  order parameter is irrelevant.
At  small anisotropy, we  observe a  first-order transition  characterized by  phase  separation. 
\end{abstract}

\maketitle

{\it Introduction.}---  
Magic-angle twisted bilayer graphene (MATBG)  provides  a  new  platform 
to    study  correlation-induced  phenomena.  Aside  from  correlated  insulating    states  
at  commensurate  fillings,  it  is of  great  present  interest  to  
understand  the   physics  
when the electron filling factor is  close  to  charge neutrality \cite{Yankowitz19}. 
 In particular,  extensive attention has  been  paid in understanding 
 how  the  correlated  insulator  gives  way  to a  superconducting state 
 via doping.      
Aside  from transport properties~\cite{Cao2018_C,Cao2018_S},  
there is a  lack of experimental tools~\cite{Zaletel_STM} to  
 study these two phases.   
 A theoretical hypothesis from Ref.~\cite{Khalaf21,Kwan_TBG}   attempts  to explain  the   two states  
 within a  unifying  framework. In  this  framework the  SU(2)  valley  symmetry  is  
 weakly  broken down  to   U(1) $\times \mathbb{Z}_2 $. 
 The  so-called Kramer inter-valley coherent insulator (K-IVC)~\cite{Bultinck20}   
 spontaneously  breaks the  U(1) charge conservation within  
 each Moir\'e valley;  skyrmion  defects of   the order parameter 
 carry electron charge $2e$, which, 
 upon  doping   condense   and   trigger   superconductivity~\cite{Khalaf21, Ippoliti22,Kwan_TBG}.    
 The success of this continuum limit picture relies crucially on the conservation of valley quantum numbers 
 corresponding to  the   \textit{chiral}  symmetry.

\begin{figure}[htbp]
\centering
\includegraphics[width=0.46\textwidth]{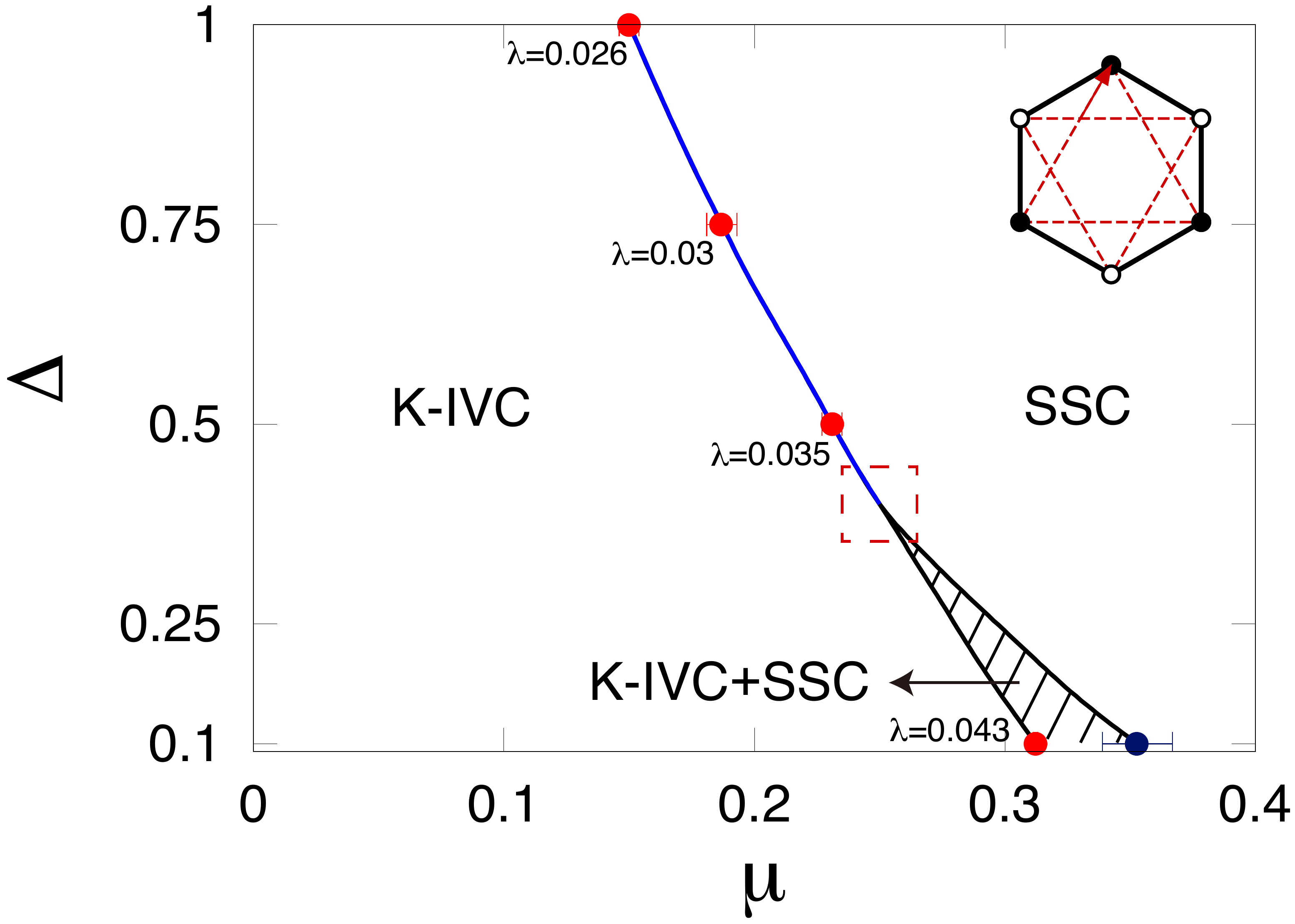}
\caption{\label{fig:Phase_diagram}
Schematic phase diagram in  the anisotropy $\Delta$ and chemical potential $\mu$ plane. Blue line indicates first-order phase transition, while black lines  continuous ones. 
For  each value of $\Delta$  we  consider a  value of $\lambda$    that maximizes the K-IVC  order  at  charge  neutrality ($\mu = 0$).
Red bullets  denote the  superconducting transition points  
while the blue dot  corresponds to the  K-IVC transition point. 
The box with dashed line indicates  the uncertainty  in  identifying  the position of critical end point, which depends on the specific choice of $\lambda$ at each value of $\Delta$.
The  inset illustrates the spin-orbit interactions inside a plaquette.}
\end{figure}

In  constructing  an  effective lattice  Hamiltonian  for  the  above,     a  major difficulty 
arises   since  the  valley  (or  chiral) symmetry  will 
 invariably be
broken by  the  regularization.  On 
the  lattice,  Dirac  cones cannot  be   rotated  independently.    Furthermore,    within  the  theory 
proposed  in Ref.~\cite{Khalaf21}   the  spin does  not  play  a  key role  in  the   pairing  mechanism,  and  in fact  spinless  versions  of 
the   theory  were  put  forward  in Ref.~\cite{Ippoliti22}.  
In this  letter   we  propose  to  \textit{swap} chiral and  spin symmetries.   We  use  the spin  degree of  freedom to  encode the (Moir\'e) valley degrees of freedom and the 
 two Moir\'e valley bands are reformulated in our case as physical spin (up and down).  
It follows naturally \cite{supplemental} that,    
the  U(1)$\times  \mathbb{Z}_2$   symmetry-broken  states  
 (K-IVC   and   Valley  Hall,  \cite{Bultinck20})  map  onto  
a dynamically generated  quantum spin Hall (QSH)
insulator     with   easy-plane   anisotropy.    
From previous works, Refs.~\cite{Grover08, Liu:2019aa, Zwang_doping},  it  is known  that 
 skyrmion excitations of the   QSH order parameter carry  charge  $2e$.   In  fact 
 in the  absence  of   easy-plane  anisotropy, it  was argued in Ref.~\cite{Zwang_doping}   
that the proliferation of  skyrmions simultaneously  destroys  the  QSH    order  and   generates superconductivity (SC) 
as the system is doped.
Hence, the low energy physics of MATBG    
can  be realized   by  including  easy-plane  anisotropy,   such that  merons, as  opposed  to  skyrmions,   become  the  low energy charged  textures.

Although    there is no essential difference between (pairs of) merons and skyrmions  in terms of  the 
defining topology and  
their associated electron charges,  the symmetry difference between U(1) $\times \mathbb{Z}_2$ and 
SU(2)  can lead to dramatic effects.  
 Depending upon  the  easy  axis anisotropy,   meron (pairs) may have a large excitation gap 
and eventually  result  in  a  doping-induced phase transition  of  mean-field     character. 
This letter aims at   
revealing the Cooper-pair condensation  when  doping a U(1)  broken  symmetry K-IVC  insulator     
and at understanding the interplay between   the  two order parameter fluctuations.

\textit{Model and Method.}--- We consider a model of Dirac fermions in $2+1$
dimensions on the honeycomb lattice with Hamiltonian   $ \hat{H} =  \hat{H}_t      +  \hat{H}_{\lambda} $.  Here, 
\begin{equation}\label{Eq:Ham_T}
\begin{aligned}
 \hat{H}_t   = - t  \sum_{ \langle \bm{i}, \bm {j} \rangle } (\hat{\ve{c}}^{\dagger}_{\bm{i} } \hat{\ve{c}}^{\phantom\dagger}_{\bm{j}}  + H.c.).
\end{aligned}
\end{equation}
The spinor
$\hat{\boldsymbol{c}}^{\dag}_{\ve{i}} =
\big(\hat{c}^{\dag}_{\ve{i},+},\hat{c}^{\dag}_{\ve{i}, - }
\big)$ where $\hat{c}^{\dag}_{\ve{i},\tau} $ creates an electron at lattice
site $\ve{i}$ with $z$-component of   the internal  degree of  freedom $\ve{\tau}$, and  the  sum  runs  over  the nearest  neighbors  of the 
honeycomb lattice. 
\begin{equation}\label{Eq:Ham_V}
\begin{aligned}
 \hat{H}_{\lambda}  = & -\lambda \sum_{\varhexagon}  \sum_{\alpha=x,y,z} \Delta^{\alpha} 
  \left( \sum_{\langle \langle \bm{i} \bm{j} \rangle \rangle  \in \varhexagon }  \hat{J}^{\alpha}_{\bm{i},\bm{j}} \right)^2,
\end{aligned}
\end{equation}  
where $ \hat{ \bm{J} }_{\bm{i},\bm{j}} \equiv i \nu_{ \bm{i} \bm{j} }
\hat{\ve{c}}^{\dagger}_{\bm{i}} \bm{\tau}
\hat{\ve{c}}^{\phantom\dagger}_{\bm{j}} + H.c.$,  with $\boldsymbol{\tau}=(\tau^x,\tau^y,\tau^z)$ the Pauli
spin matrices,  see the inset of Fig. \ref{fig:Phase_diagram}. This term is a plaquette
interaction involving next-nearest-neighbor pairs of sites and phase factors
$\nu_{\boldsymbol{ij}}=\pm1$ identical to the Kane-Mele model
\cite{KaneMele05b}, see also Ref.~\onlinecite{Liu:2019aa}.  Finally,   the   
easy-plane  anisotropy  is  imposed  by  choosing $\ve{\Delta} =  \left(  1,1, \Delta  \right)$. 

The SU(2) invariant version ($\Delta=1$) of this Hamiltonian has been well studied 
in Ref.~\onlinecite{Liu:2019aa,Zwang_doping}. 
Associating $\tau$ to the spin  degree of  freedom,   a   dynamically generated QSH insulator that 
spontaneously breaks SU(2) $\tau$-rotational symmetry  
is found at intermediate  interacting strength ($\lambda$) and  at half-filling. An SSC can be 
realized by increasing $\lambda$ or doping. For $\Delta<1$, 
we realized a U(1) symmetry-broken QSH state by reducing the full $\tau$-rotational symmetry of the 
Hamiltonian \cite{EP_QSH}.

The Hamiltonian we considered here captures the key ingredients of MATBG around charge neutrality.  
In MATBG, the Chern number of the flat bands is associated with   
 the (Moir\'e) valley quantum number~\cite{Bultinck20}.     
 As documented in the supplemental material, the spin degrees of freedom in our toy model play the role of the (Moir\'e) 
 valley in their case.  
Therefore, the Kramers inter-valley coherent phase, which breaks spontaneously the  U(1) valley symmetry is not different than our QSH insulator from a symmetry point of view.   
The following consequence from this symmetry argument is that in both cases, electron pairing arises from Kramers doublet based on pairs of meron configurations in the U(1) broken insulator.  
Upon doping, these pre-formed pairs condensate and superconductivity is formed.    
Crucially, although charge conservation within each valley (e.g., chiral symmetry) is an exact symmetry in the continuum limit, it is generically not possible  realize it in a lattice Hamiltonian.   
A simple way of performing this regularization is to substitute the valley degrees of freedom with spin ones  
~\footnote{From now on, we identify the K-IVC state in continuum description of MATBG as the 
dynamically generated QSH state in our model, and keep the name of K-IVC through the text. }.

Here we  focus on the case  where $ \Delta \in [ 0, 1 ) $
such that  the SU(2) spin rotational symmetry is reduced to U(1)$\times  \mathbb{Z}_2$. 
We investigate three values of the  anisotropy and only consider   
values of $\lambda$ deep inside the charge-neutral K-IVC state \cite{EP_QSH}:
$\lambda=0.043$ for $\Delta=0.1$, $\lambda=0.035$ for $\Delta=0.5$, and $\lambda=0.03$ for $\Delta=0.75$.  
Our  unit of energy  is  set  by $t=1$.

We used the ALF (Algorithms for Lattice Fermions)
implementation~\cite{ALF_v2} of the auxiliary-field quantum
Monte Carlo (QMC) method~\cite{Blankenbecler81,White89,Assaad08_rev}.  Because
$\lambda > 0$ and $\Delta>0$,  a real Hubbard-Stratonovich decomposition for the
perfect square term does not break the time-reversal symmetry.    Since  charge  is  
conserved,  the  eigenvalues of  the  fermion  determinant  come  in  
complex  conjugate  pairs  such  that no sign problem  occurs  \cite{Wu04}.
We simulate periodic systems with size of $L\times L$. 
The imaginary time interval is $\Delta \tau=0.2$ and  a symmetric Trotter decomposition is chosen  
to ensure the  hermiticity of  the   Trotterized     imaginary  time evolution \cite{Liu:2019aa}. 
Additionally, we apply a checkerboard decomposition  to the exponential of hopping matrix $\hat{H}_t$.
Following our previous work~\cite{Zwang_doping},  we used a
projective QMC algorithm (PQMC)
\cite{Sugiyama86,Sorella89,Assaad08_rev}.
 For  the trial  wave  function, we take  the ground state of  the  hopping Hamiltonian in  Eq.~(\ref{Eq:Ham_T}) with spin-dependent twisted boundary  conditions.   This Slater determinant is then uniquely defined and  
 also  time-reversal symmetric.

{\it QMC results.}--- 
To capture the K-IVC order, we define the local   operators   
$ \hat{\bm{O}}_{ \ve{r}, n } \equiv 
\hat{ \bm{J} }_{\ve{r} +
\ve{\delta}_n,\ve{r} + \ve{\eta}_n}$. 
Here,  $\ve{r}$ denotes  a unit cell    
and $n=1,2,...6$ are the six next-nearest neighbor bonds of the corresponding
hexagon with legs $\ve{r} + \ve{\delta}_n$ and $\ve{r} + \ve{\eta}_n$.   
The K-IVC order with broken U(1)  rotational symmetry 
 is detected by  computing: 
\begin{equation}
\begin{aligned}
   S^{\text{K-IVC} }_{m, n} (\boldsymbol{q})
        \equiv &  \frac{1}{L^2} \sum_{\boldsymbol{r},\boldsymbol{r'}}    e^{\mathrm{i}\boldsymbol{q}\cdot(\boldsymbol{r}-\boldsymbol{r}')} 
        \langle  \hat{ O }^X_{\boldsymbol{r}, m} \hat{  O  }^X_{\boldsymbol{r'}, n }
    +
     \hat{ O }^Y_{\boldsymbol{r},  m}  \hat{ O }^Y_{\boldsymbol{r'},  n}
    \rangle,
\end{aligned}
\end{equation}
with $m,n=1,2,...,6$.

\begin{figure}[htbp]
\centering
\includegraphics[width=0.4\textwidth]{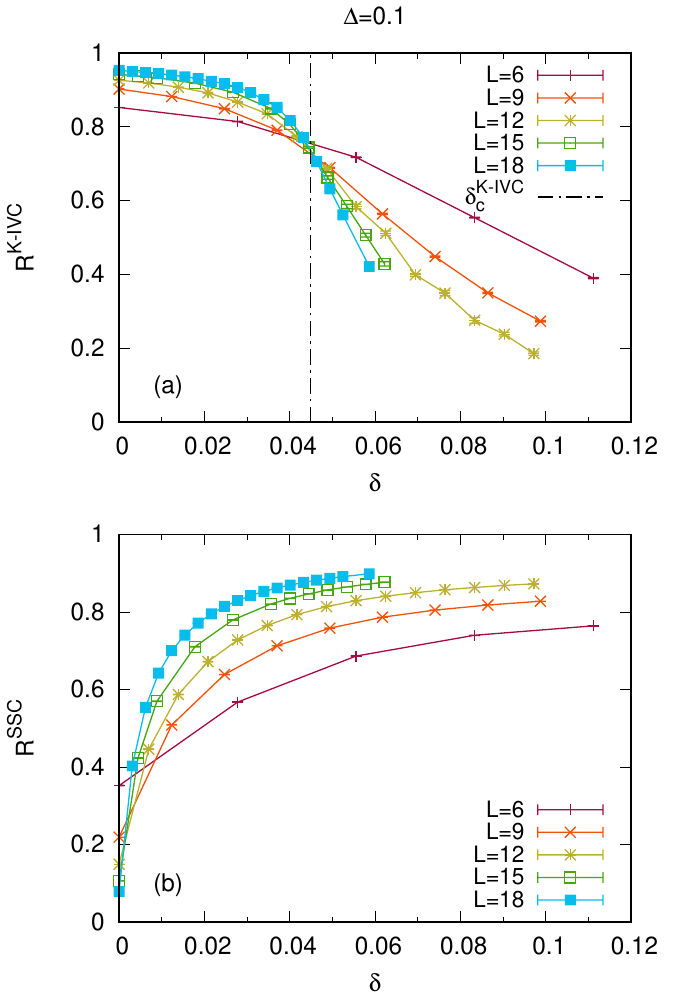}
\caption{\label{fig:Ratioeq_0.1}
Equal-time correlation ratio for (a) $R^{\rm K-IVC}$ and (b) $R^{\rm SSC}$ as a function of doping factor $\delta$ for $\Delta=0.1$. 
The vertical dashed line is a guide to the eye, fitted from the crossing point of K-IVC order parameter 
 between $L=15$ and $18$.  
}
\end{figure}

\begin{figure}[htbp]
\centering
\includegraphics[width=0.4\textwidth]{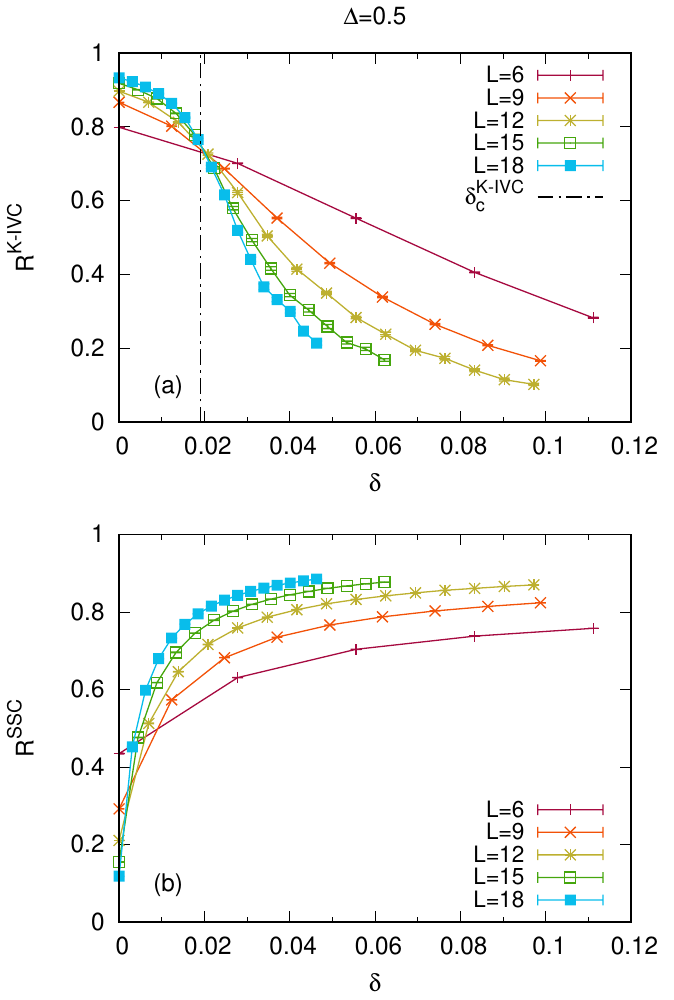}
\caption{\label{fig:Ratioeq_0.5}
Same as Fig.~\ref{fig:Ratioeq_0.1}, for $\Delta=0.5$.
}
\end{figure}

As  for  SSC   we  consider:
\begin{equation} 
\begin{aligned}
   S^{\text{SSC} }_{a, b} (\boldsymbol{q})
        \equiv  &  \frac{1}{L^2} \sum_{\boldsymbol{r},\boldsymbol{r'}}     e^{\mathrm{i}\boldsymbol{q}\cdot(\boldsymbol{r}-\boldsymbol{r}')}     
         [ \langle  \hat{ \eta }^{+}_{\boldsymbol{r},\boldsymbol{ \widetilde{\delta} }_a } \hat{ \eta  }^{-}_{\boldsymbol{r'},\boldsymbol{ \widetilde{ \delta} }_b }  \rangle +
         \langle  \hat{ \eta }^{- }_{\boldsymbol{r},\boldsymbol{ \widetilde{\delta} }_a }  \hat{ \eta  }^{+}_{\boldsymbol{r'},\boldsymbol{ \widetilde{\delta} }_b }
        \rangle],
\end{aligned}
\end{equation}
where $a,b=1,2$, denotes the A(B) sublattices, and
$  \hat{\eta}^{+}_{\ve{r},\ve{\tilde{\delta}_a}} = \hat{c}^{\dagger}_{\ve{r} 
  +\ve{\tilde{\delta}_a}, + }  \hat{c}^{\dagger}_{\ve{r}
 +\ve{\tilde{\delta}_a},-}  $.

The corresponding renormalization-gourp invariant correlation ratio reads: 
\begin{equation}
  R^{O} \equiv  1 - \frac{S^O(\bm{q}_{0}+\delta \bm{q})}{S^O(\bm{q}_{0}) },
\end{equation}
where $S^O$ is the largest eigenvalue of the corresponding correlation matrix ($O$=K-IVC,SSC). 
$\bm{q}_{0}$ is the ordering wave vector and $\bm{q}_{0}+\delta \bm{q}$ is the neighboring wave vector
($|\delta \bm{q}|=\frac{4\pi}{\sqrt{3}L}$).

We define the doping factor $\delta$ by the density of doped holes relative to charge neutrality,  
$ \delta \equiv  \frac{2L^2 -   N_e }{2 L^2}$.   Here,  $ N_e$   counts  the  number of electrons. 
Due to  large  auto-correlation times  related  to  particle  fluctuations,  it  is  convenient  to  adopt 
a  canonical  ensemble.   Within this ensemble,  
the  chemical potential is   
evaluated from the slope of the `tower of states' in  the charge sector:   
\begin{equation}
  \begin{aligned}
   \mu  \equiv  \frac{\Delta_{ \eta^{-} }(N_\text{e})}{2}\,.
  \end{aligned}
   \label{Eq:cp-pqmc}
\end{equation}  
Here,  $ \Delta_{ \eta^{-} }(N_\text{e})$ is the  pairing ($\eta^{-}$) gap extrapolated from
the time-displaced correlation function.

Our  results  are  summarized in the  ground state phase diagram in  the  $\Delta$ versus 
$\mu$   plane,  as shown in  Fig.~\ref{fig:Phase_diagram}.        For   each  value of  $\Delta$  we consider  
a  coupling $\lambda$   that  places  us well   within the   correlated  insulating  phase  at   charge   neutrality \cite{EP_QSH}.
Below  we will document  that  at  small  anisotropy  we   observe  a  first-order  transition   characterized    by  
phase  separation between  the  K-IVC and  SSC.   At larger  values of $\Delta$  phase  coexistence  sets  in.  

\begin{figure}[htbp]
\centering
\includegraphics[width=0.4\textwidth]{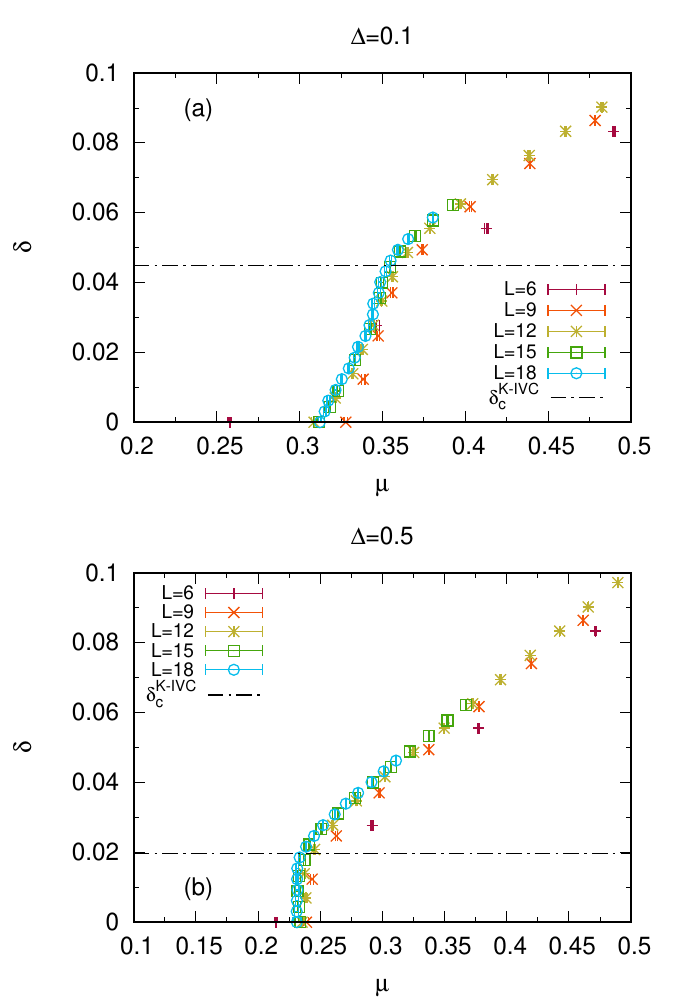}
\caption{\label{fig:chem}
Doping factor $\delta$ as a function of chemical potential $\mu$ 
for (a) $\Delta=0.1$ and (b) $\Delta=0.5$. 
The horizontal line is a guide to the eye, which is the same as the vertical one in Figs.~\ref{fig:Ratioeq_0.1} and \ref{fig:Ratioeq_0.5}.  
}
\end{figure}

Figures~\ref{fig:Ratioeq_0.1}  and  \ref{fig:Ratioeq_0.5}    show  the correlation ratios  for  the  SSC  and  K-IVC  instabilities   
at  $\Delta= 0.1$  and  $\Delta = 0.5$.    
As apparent,    the  K-IVC  order  survives finite doping,  with  the  critical  doping,  
$\delta^{ \text{ K-IVC}}_c$,  being given by  the   
crossing point of the  curves.  On  the other  hand,   superconductivity  sets  in   at  any  finite  doping  for  both considered 
 values of  the  anisotropy,  $\Delta$.     Since  we  are   working in  a canonical ensemble,   we  have  to  check  for   
  the  stability    against   phase  separation as signaled by  an  infinite  compressibility,  $\frac{\partial \delta}{\partial \mu} $. 
   Figure~\ref{fig:chem}(b)  plots  this  quantity  at, $\Delta = 0.5$.   As apparent,  in the  range  
   $0  <  \delta   < \delta^{\text{ K-IVC}}_c $   the 
   data is  consistent  with a  diverging  compressibility.       Hence,   in this    doping  range and in the thermodynamic limit,  
  we expect  to observe  puddles   of  K-IVC   insulating phases    with  a total  density   set   by 
  $  1 -  \frac{\delta}{\delta^{\text{K-IVC} }_c} $  and  
   regions of  SSC with a total  density   set   by   $ \frac{\delta}{\delta^{\text{K-IVC}}_c} $  \footnote{ For a finite size calculation  in  the 
 canonical ensemble, one may need  a very  large system  to observe  phase  separation due  to  the energetics  of  the  domain boundary.}.

On the other hand,   at   strong  anisotropy,   $\Delta=0.1$,   Fig.~\ref{fig:chem}(a) shows  no  sign of  diverging  
compressibility,   thus  indicating   phase  coexistence.  
However  we  observe  two   non-analytical  points,  at $\mu_{c1} \approx 0.31 $  
and $ \mu_{c2} \approx 0.35$  corresponds  to  the   insulator   to  superconductor  transition   
at  $\delta  = 0^{+}$  and  to  the  K-IVC  transition at  $\delta =  \delta^{\text{K-IVC}}_c$. 
Both  transitions    show  no  features  of   first-order  transitions, as can be directly seen  from  
the continuous behavior of $\delta-\mu$ dependence in Fig.~\ref{fig:chem}(a).   
We  note  that at the  superconducting transition at $\mu_{c1}$,   the hyper-scaling law:    
\begin{equation}
  \delta - \delta_c   \propto  | \mu - \mu_c |^{ \nu d }  , \quad 
  \nu z = 1      
\end{equation}  
holds. 
This is due to the fact that  the generator of the SSC order parameter  
couples to the chemical potential, $\mu$,  corresponding  to  
the tuning parameter of this quantum phase transition  \cite{Fisher1989}. 

The linear $\delta$-$\mu$ dependence at  $\mu_{c1} \approx 0.31$ 
in Fig.~\ref{fig:chem}(a)  suggests that $z=2$ at the superconducting phase transition.   
   Assuming that the background of K-IVC ordering does not couple to any critical fluctuations
at the superconducting critical point,  a mean field phase transition with $z=2, \nu=0.5$  
at $\mu_{c1}$  is   expected.   
This transition is in  the  very   same universality  class  as   that of  the  Mott-insulator-superfluid 
transition  in  doped Bose-Hubbard system~\cite{Fisher1989}.

\begin{figure}[htbp]
\centering
\includegraphics[width=0.4\textwidth]{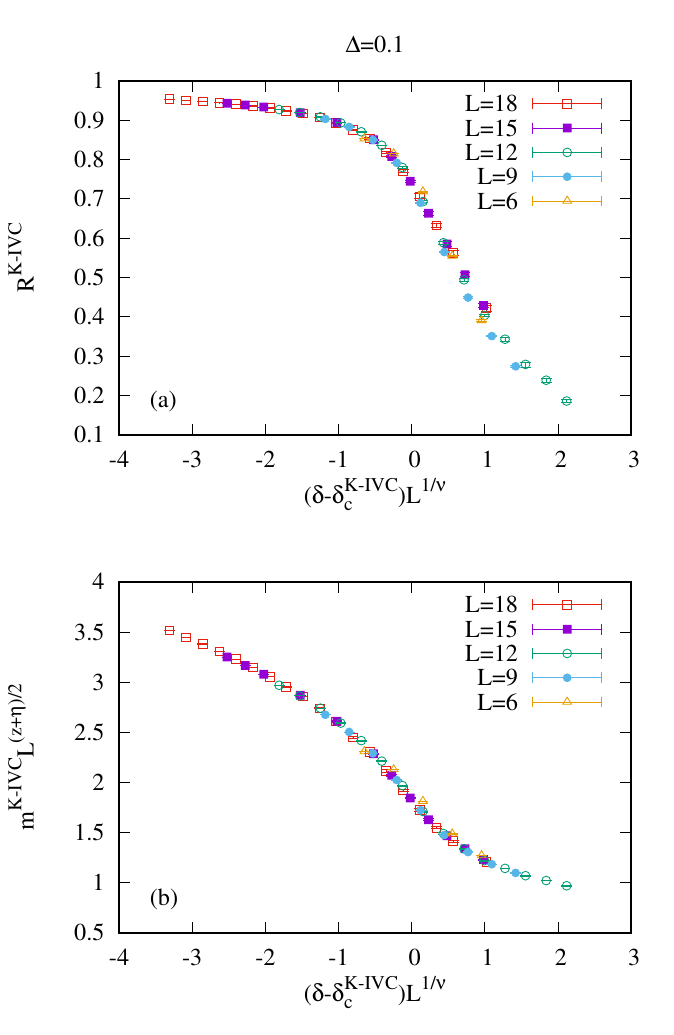}
\caption{\label{fig:colla1}
Data collapse at $\Delta=0.1$ with $\delta_c^{\rm K-IVC} = 0.0448$ and 3D XY exponent whose $\nu = 0.67169(7)$, $\eta=0.03810(8)$ for (a) the correlation ratio and (b) the K-IVC order parameter. 
}
\end{figure}

To  understand  the  nature of  the   K-IVC transition  in the  background  of  superconducting order,  we  fit  our   data  to  the following form: 
\begin{equation}\label{Eq:fit}
\begin{aligned}
 & R^{\text{K-IVC}} = f_1 ( (\delta - \delta^{\text{K-IVC}}_c ) L^{ 1 / \nu } ),  \\ 
 & m^{\text{K-IVC}} L^{ ( z + \eta )/2 } =  f_2 ( (\delta - \delta^{\text{K-IVC}}_c ) L^{ 1 / \nu } ).
\end{aligned}
\end{equation}
where $  m^{\text{K-IVC}} \equiv  \sqrt{ \sum_n S^{\text{K-IVC}}_{n, n} (\bm{q}_0 ) /L^2 } $ is the K-IVC order parameter. 
Our  results,  Fig.~\ref{fig:colla1},   suggest that  the  transition  is  consistent  with  Lorentz invariance,  $z=1$,   and 
that it falls  into  the 3D XY universality class.   
In particular   we   considered  $ \nu \approx 0.67169 $ and $ \eta \approx 0.03810  $ from previous Monte Carlo 
simulations of  the  3D XY model \cite{Hasenbusch2019} as  well as  $\delta^{\text{K-IVC}}_c=0.0448$ for our data collapse.     
As shown in Fig.~\ref{fig:colla1},  both two quantities show nice collapse 
for system sizes from $L=6$ to $L=18$.

\begin{table}[htbp]
\caption{\label{tab:nu}
    $\nu$ fitting of K-IVC vanishing transition }
	\centering
	\begin{tabular}{ccccccc}
		\toprule 
		\hline
		order & $L$ & $(\delta-\delta^{\text{K-IVC}}_c)L^{1/\nu}$  & $\nu$ & $\chi_r^2$/DOF &  observable \\ 
		\midrule  
		\hline
		13
   &[15:18] & [-2.8:1.05]  & 0.67(1)  & 1.19/16 & $R^{\rm K-IVC}$ \\
		7  &[15:18] & [-2.58:1.05] & 0.67(1)  & 1.39/15 & $m^{\rm K-IVC}$ \\
		\hline
        \bottomrule 
	\end{tabular}
\end{table}

We also performed a collective polynomial fit  using $L=15$ and $18$ based on Eq.~(\ref{Eq:fit}).
The results are listed in Tab.~\ref{tab:nu}.  
Even when a wide fitting range is taken into consideration,   
the value of $\nu$ that we obtain is consistent with 3D XY universality class.

An interesting question to ask is why  symmetry allowed coupling terms between  the K-IVC  order parameter and 
the critical charge fluctuations of   the  superconductor play  no role at the 3D XY phase transition. 
In   Ref.~\cite{supplemental}  we  provide   a  power  counting  argument to show  that the  Goldstone
modes of  the   SSC   order  parameter  are  irrelevant  at  the    3D  XY     fixed  point.

{\it Discussion and summary.}---\  
We  have  considered  an  effective  lattice  model  that  captures  the  physics  of  a  candidate  theory  of  
MATBG \cite{Khalaf21},  which unifies  the   K-IVC   insulator  and   superconducting phases:   skyrmions  defect
of  the  K-IVC  order parameter    carry   charge  $2e$   and  condense  upon  doping. The  key  insight  to  obtaining  
lattice  regularization of this  physics  is  to encode  the valley  symmetry   with the spin degree of  freedom.  
In Ref.~\cite{supplemental} we  show that the  single-particle  gap  does not vanish.  This  allows  us to  
integrate out the electronic degrees of  freedom   and precisely obtain the   low  energy  topological  
field  theory  presented  in Ref.~\cite{Khalaf21}.
Large-scale QMC simulations 
reveal   two different types of  transitions    depending  upon  the strength of the easy-plane anisotropy. 
 At  large  anisotropy   we  observe  phase  coexistence.     As  a  function of  doping    the  
  insulating  state  gives  way  to a  superconducting  phase,   with the  universality  class  being  identical  
 to  that  of  the  Bose-Hubbard model~\cite{Fisher1989}.   At  larger  doping  we  observe  the  
 vanishing of  the    U(1)   order in the  background of the superconducting phase.     Our results show
 that this   phase  transition belongs  to  the  3D XY   universality class,  such  that   coupling  to   
 gapless   charge  fluctuations  are  irrelevant  at  this critical point.

For  small   anisotropic case, we observe phase separation.     In    conjunction  with our  previous  results
for  the SU(2)  case~\cite{Zwang_doping},    we    observe  that the  doping  range  in  which   phase  separation 
occurs      decreases   as  one  approaches  the SU(2)  symmetric  point,   $\Delta \rightarrow 1$.  
This challenges the  conclusion  of  Ref.~\cite{Zwang_doping} where a 
seemingly continuous transition with large dynamical exponent is observed. 
Alternatively, if symmetry plays a crucial role, an interesting possibility is that  
the large $z$ continuous transition only exists in the  SU(2) symmetric case.  
 Our  parameters are  such that  the  single-particle gap  
at  charge  neutrality  is  independent of the anisotropy and   the  pairing gap   decreases as the  anisotropy   grows  \cite{supplemental}.  
 We  interpret this 
in  terms of  merons pairs   that  become  energetically 
more  expensive due  to  a  smaller  core size  at  strong anisotropy.    It is  hence  tempting 
to  interpret  the phase  diagram of  Fig.~\ref{fig:Phase_diagram}   as  dominated  by  topology in the vicinity of the 
SU(2)  symmetric  point   and   from the point of  view of a   Ginzburg-Landau   theory  at  strong  anisotropy.   
This  statement is  substantiated  by  calculations in Ref.~\cite{supplemental}  at  small  anisotropy,  that show  the  locking of   the 
charge  density  and curvature  of the   K-IVC  order parameter.

   Let us  now  return  to   MATBG,   where  the   easy-plane   anisotropy  is  expected  to  be small
 \cite{Bultinck20}.   The numerical study in Ref.~\cite{Ippoliti22} suggests that the mechanism of skyrmion superconductivity is  stable to 
 the  Coulomb repulsion, such that our  model can very  well    capture  the low energy physics  of   MATBG.
In this case,   our  results  suggest  that    the   doping-induced  transition  is 
 first-order and   entails   phase  separation.

Z.W. thanks  for enlightening conversations with Benoit Doucot and Peng Rao.     
The authors gratefully acknowledge the Gauss Centre for Supercomputing e.V. for funding this
project by providing computing time on the GCS Supercomputer SUPERMUC-NG at Leibniz Supercomputing Centre.
F.F.A. acknowledges the  DFG  for   funding  via  W\"urzburg-Dresden Cluster of 
Excellence on Complexity and Topology in Quantum Matter
ct.qmat (EXC 2147, Project ID 390858490)  as  well  as 
the SFB1170 on Topological and Correlated Electronics at Surfaces and Interfaces.
D.H. and W.G. were
supported by the National Natural Science Foundation of
China under Grants No. 12175015 and No. 11734002.
T.S. acknowledges funding from the Deutsche Forschungsgemeinschaft under Grant No. SA 3986/1-1.
Y.L. was supported by the National Natural Science Foundation of China under Grant No.~11947232 and the Fundamental Research Funds for the Central Universities from the Beijing University of Posts and Telecommunications under Grant No.~2023RC42. 
Z.W. was supported by the FP7/ERC Consolidator Grant QSIMCORR, No. 771891. 

\bibliography{ref,fassaad}

\begin{thebibliography}{29}%
\makeatletter
\providecommand \@ifxundefined [1]{%
 \@ifx{#1\undefined}
}%
\providecommand \@ifnum [1]{%
 \ifnum #1\expandafter \@firstoftwo
 \else \expandafter \@secondoftwo
 \fi
}%
\providecommand \@ifx [1]{%
 \ifx #1\expandafter \@firstoftwo
 \else \expandafter \@secondoftwo
 \fi
}%
\providecommand \natexlab [1]{#1}%
\providecommand \enquote  [1]{``#1''}%
\providecommand \bibnamefont  [1]{#1}%
\providecommand \bibfnamefont [1]{#1}%
\providecommand \citenamefont [1]{#1}%
\providecommand \href@noop [0]{\@secondoftwo}%
\providecommand \href [0]{\begingroup \@sanitize@url \@href}%
\providecommand \@href[1]{\@@startlink{#1}\@@href}%
\providecommand \@@href[1]{\endgroup#1\@@endlink}%
\providecommand \@sanitize@url [0]{\catcode `\\12\catcode `\$12\catcode
  `\&12\catcode `\#12\catcode `\^12\catcode `\_12\catcode `\%12\relax}%
\providecommand \@@startlink[1]{}%
\providecommand \@@endlink[0]{}%
\providecommand \url  [0]{\begingroup\@sanitize@url \@url }%
\providecommand \@url [1]{\endgroup\@href {#1}{\urlprefix }}%
\providecommand \urlprefix  [0]{URL }%
\providecommand \Eprint [0]{\href }%
\providecommand \doibase [0]{http://dx.doi.org/}%
\providecommand \selectlanguage [0]{\@gobble}%
\providecommand \bibinfo  [0]{\@secondoftwo}%
\providecommand \bibfield  [0]{\@secondoftwo}%
\providecommand \translation [1]{[#1]}%
\providecommand \BibitemOpen [0]{}%
\providecommand \bibitemStop [0]{}%
\providecommand \bibitemNoStop [0]{.\EOS\space}%
\providecommand \EOS [0]{\spacefactor3000\relax}%
\providecommand \BibitemShut  [1]{\csname bibitem#1\endcsname}%
\let\auto@bib@innerbib\@empty
\bibitem [{\citenamefont {Yankowitz}\ \emph {et~al.}(2019)\citenamefont
  {Yankowitz}, \citenamefont {Chen}, \citenamefont {Polshyn}, \citenamefont
  {Zhang}, \citenamefont {Watanabe}, \citenamefont {Taniguchi}, \citenamefont
  {Graf}, \citenamefont {Young},\ and\ \citenamefont {Dean}}]{Yankowitz19}%
  \BibitemOpen
  \bibfield  {author} {\bibinfo {author} {\bibfnamefont {M.}~\bibnamefont
  {Yankowitz}}, \bibinfo {author} {\bibfnamefont {S.}~\bibnamefont {Chen}},
  \bibinfo {author} {\bibfnamefont {H.}~\bibnamefont {Polshyn}}, \bibinfo
  {author} {\bibfnamefont {Y.}~\bibnamefont {Zhang}}, \bibinfo {author}
  {\bibfnamefont {K.}~\bibnamefont {Watanabe}}, \bibinfo {author}
  {\bibfnamefont {T.}~\bibnamefont {Taniguchi}}, \bibinfo {author}
  {\bibfnamefont {D.}~\bibnamefont {Graf}}, \bibinfo {author} {\bibfnamefont
  {A.~F.}\ \bibnamefont {Young}}, \ and\ \bibinfo {author} {\bibfnamefont
  {C.~R.}\ \bibnamefont {Dean}},\ }\href {\doibase 10.1126/science.aav1910}
  {\bibfield  {journal} {\bibinfo  {journal} {Science}\ }\textbf {\bibinfo
  {volume} {363}},\ \bibinfo {pages} {1059} (\bibinfo {year} {2019})},\ \Eprint
  {http://arxiv.org/abs/https://www.science.org/doi/pdf/10.1126/science.aav1910}
  {https://www.science.org/doi/pdf/10.1126/science.aav1910} \BibitemShut
  {NoStop}%
\bibitem [{\citenamefont {Cao}\ \emph {et~al.}(2018{\natexlab{a}})\citenamefont
  {Cao}, \citenamefont {Fatemi}, \citenamefont {Demir}, \citenamefont {Fang},
  \citenamefont {Tomarken}, \citenamefont {Luo}, \citenamefont
  {Sanchez-Yamagishi}, \citenamefont {Watanabe}, \citenamefont {Taniguchi},
  \citenamefont {Kaxiras}, \citenamefont {Ashoori},\ and\ \citenamefont
  {Jarillo-Herrero}}]{Cao2018_C}%
  \BibitemOpen
  \bibfield  {author} {\bibinfo {author} {\bibfnamefont {Y.}~\bibnamefont
  {Cao}}, \bibinfo {author} {\bibfnamefont {V.}~\bibnamefont {Fatemi}},
  \bibinfo {author} {\bibfnamefont {A.}~\bibnamefont {Demir}}, \bibinfo
  {author} {\bibfnamefont {S.}~\bibnamefont {Fang}}, \bibinfo {author}
  {\bibfnamefont {S.~L.}\ \bibnamefont {Tomarken}}, \bibinfo {author}
  {\bibfnamefont {J.~Y.}\ \bibnamefont {Luo}}, \bibinfo {author} {\bibfnamefont
  {J.~D.}\ \bibnamefont {Sanchez-Yamagishi}}, \bibinfo {author} {\bibfnamefont
  {K.}~\bibnamefont {Watanabe}}, \bibinfo {author} {\bibfnamefont
  {T.}~\bibnamefont {Taniguchi}}, \bibinfo {author} {\bibfnamefont
  {E.}~\bibnamefont {Kaxiras}}, \bibinfo {author} {\bibfnamefont {R.~C.}\
  \bibnamefont {Ashoori}}, \ and\ \bibinfo {author} {\bibfnamefont
  {P.}~\bibnamefont {Jarillo-Herrero}},\ }\href {\doibase 10.1038/nature26154}
  {\bibfield  {journal} {\bibinfo  {journal} {Nature}\ }\textbf {\bibinfo
  {volume} {556}},\ \bibinfo {pages} {80} (\bibinfo {year}
  {2018}{\natexlab{a}})}\BibitemShut {NoStop}%
\bibitem [{\citenamefont {Cao}\ \emph {et~al.}(2018{\natexlab{b}})\citenamefont
  {Cao}, \citenamefont {Fatemi}, \citenamefont {Fang}, \citenamefont
  {Watanabe}, \citenamefont {Taniguchi}, \citenamefont {Kaxiras},\ and\
  \citenamefont {Jarillo-Herrero}}]{Cao2018_S}%
  \BibitemOpen
  \bibfield  {author} {\bibinfo {author} {\bibfnamefont {Y.}~\bibnamefont
  {Cao}}, \bibinfo {author} {\bibfnamefont {V.}~\bibnamefont {Fatemi}},
  \bibinfo {author} {\bibfnamefont {S.}~\bibnamefont {Fang}}, \bibinfo {author}
  {\bibfnamefont {K.}~\bibnamefont {Watanabe}}, \bibinfo {author}
  {\bibfnamefont {T.}~\bibnamefont {Taniguchi}}, \bibinfo {author}
  {\bibfnamefont {E.}~\bibnamefont {Kaxiras}}, \ and\ \bibinfo {author}
  {\bibfnamefont {P.}~\bibnamefont {Jarillo-Herrero}},\ }\href {\doibase
  10.1038/nature26160} {\bibfield  {journal} {\bibinfo  {journal} {Nature}\
  }\textbf {\bibinfo {volume} {556}},\ \bibinfo {pages} {43} (\bibinfo {year}
  {2018}{\natexlab{b}})}\BibitemShut {NoStop}%
\bibitem [{\citenamefont {Hong}\ \emph {et~al.}(2022)\citenamefont {Hong},
  \citenamefont {Soejima},\ and\ \citenamefont {Zaletel}}]{Zaletel_STM}%
  \BibitemOpen
  \bibfield  {author} {\bibinfo {author} {\bibfnamefont {J.~P.}\ \bibnamefont
  {Hong}}, \bibinfo {author} {\bibfnamefont {T.}~\bibnamefont {Soejima}}, \
  and\ \bibinfo {author} {\bibfnamefont {M.~P.}\ \bibnamefont {Zaletel}},\
  }\href {\doibase 10.1103/PhysRevLett.129.147001} {\bibfield  {journal}
  {\bibinfo  {journal} {Phys. Rev. Lett.}\ }\textbf {\bibinfo {volume} {129}},\
  \bibinfo {pages} {147001} (\bibinfo {year} {2022})}\BibitemShut {NoStop}%
\bibitem [{\citenamefont {Khalaf}\ \emph {et~al.}(2021)\citenamefont {Khalaf},
  \citenamefont {Chatterjee}, \citenamefont {Bultinck}, \citenamefont
  {Zaletel},\ and\ \citenamefont {Vishwanath}}]{Khalaf21}%
  \BibitemOpen
  \bibfield  {author} {\bibinfo {author} {\bibfnamefont {E.}~\bibnamefont
  {Khalaf}}, \bibinfo {author} {\bibfnamefont {S.}~\bibnamefont {Chatterjee}},
  \bibinfo {author} {\bibfnamefont {N.}~\bibnamefont {Bultinck}}, \bibinfo
  {author} {\bibfnamefont {M.~P.}\ \bibnamefont {Zaletel}}, \ and\ \bibinfo
  {author} {\bibfnamefont {A.}~\bibnamefont {Vishwanath}},\ }\href {\doibase
  10.1126/sciadv.abf5299} {\bibfield  {journal} {\bibinfo  {journal} {Science
  Advances}\ }\textbf {\bibinfo {volume} {7}},\ \bibinfo {pages} {eabf5299}
  (\bibinfo {year} {2021})},\ \Eprint
  {http://arxiv.org/abs/https://www.science.org/doi/pdf/10.1126/sciadv.abf5299}
  {https://www.science.org/doi/pdf/10.1126/sciadv.abf5299} \BibitemShut
  {NoStop}%
\bibitem [{\citenamefont {Kwan}\ \emph {et~al.}(2022)\citenamefont {Kwan},
  \citenamefont {Wagner}, \citenamefont {Bultinck}, \citenamefont {Simon},\
  and\ \citenamefont {Parameswaran}}]{Kwan_TBG}%
  \BibitemOpen
  \bibfield  {author} {\bibinfo {author} {\bibfnamefont {Y.~H.}\ \bibnamefont
  {Kwan}}, \bibinfo {author} {\bibfnamefont {G.}~\bibnamefont {Wagner}},
  \bibinfo {author} {\bibfnamefont {N.}~\bibnamefont {Bultinck}}, \bibinfo
  {author} {\bibfnamefont {S.~H.}\ \bibnamefont {Simon}}, \ and\ \bibinfo
  {author} {\bibfnamefont {S.~A.}\ \bibnamefont {Parameswaran}},\ }\href
  {\doibase 10.1103/PhysRevX.12.031020} {\bibfield  {journal} {\bibinfo
  {journal} {Phys. Rev. X}\ }\textbf {\bibinfo {volume} {12}},\ \bibinfo
  {pages} {031020} (\bibinfo {year} {2022})}\BibitemShut {NoStop}%
\bibitem [{\citenamefont {Bultinck}\ \emph {et~al.}(2020)\citenamefont
  {Bultinck}, \citenamefont {Khalaf}, \citenamefont {Liu}, \citenamefont
  {Chatterjee}, \citenamefont {Vishwanath},\ and\ \citenamefont
  {Zaletel}}]{Bultinck20}%
  \BibitemOpen
  \bibfield  {author} {\bibinfo {author} {\bibfnamefont {N.}~\bibnamefont
  {Bultinck}}, \bibinfo {author} {\bibfnamefont {E.}~\bibnamefont {Khalaf}},
  \bibinfo {author} {\bibfnamefont {S.}~\bibnamefont {Liu}}, \bibinfo {author}
  {\bibfnamefont {S.}~\bibnamefont {Chatterjee}}, \bibinfo {author}
  {\bibfnamefont {A.}~\bibnamefont {Vishwanath}}, \ and\ \bibinfo {author}
  {\bibfnamefont {M.~P.}\ \bibnamefont {Zaletel}},\ }\href {\doibase
  10.1103/PhysRevX.10.031034} {\bibfield  {journal} {\bibinfo  {journal} {Phys.
  Rev. X}\ }\textbf {\bibinfo {volume} {10}},\ \bibinfo {pages} {031034}
  (\bibinfo {year} {2020})}\BibitemShut {NoStop}%
\bibitem [{\citenamefont {Chatterjee}\ \emph {et~al.}(2022)\citenamefont
  {Chatterjee}, \citenamefont {Ippoliti},\ and\ \citenamefont
  {Zaletel}}]{Ippoliti22}%
  \BibitemOpen
  \bibfield  {author} {\bibinfo {author} {\bibfnamefont {S.}~\bibnamefont
  {Chatterjee}}, \bibinfo {author} {\bibfnamefont {M.}~\bibnamefont
  {Ippoliti}}, \ and\ \bibinfo {author} {\bibfnamefont {M.~P.}\ \bibnamefont
  {Zaletel}},\ }\href {\doibase 10.1103/PhysRevB.106.035421} {\bibfield
  {journal} {\bibinfo  {journal} {Phys. Rev. B}\ }\textbf {\bibinfo {volume}
  {106}},\ \bibinfo {pages} {035421} (\bibinfo {year} {2022})}\BibitemShut
  {NoStop}%
\bibitem [{sup()}]{supplemental}%
  \BibitemOpen
  \href@noop {} {}\bibinfo {note} {See Supplemental Material.}\BibitemShut
  {Stop}%
\bibitem [{\citenamefont {Grover}\ and\ \citenamefont
  {Senthil}(2008)}]{Grover08}%
  \BibitemOpen
  \bibfield  {author} {\bibinfo {author} {\bibfnamefont {T.}~\bibnamefont
  {Grover}}\ and\ \bibinfo {author} {\bibfnamefont {T.}~\bibnamefont
  {Senthil}},\ }\href {\doibase 10.1103/PhysRevLett.100.156804} {\bibfield
  {journal} {\bibinfo  {journal} {Phys. Rev. Lett.}\ }\textbf {\bibinfo
  {volume} {100}},\ \bibinfo {pages} {156804} (\bibinfo {year}
  {2008})}\BibitemShut {NoStop}%
\bibitem [{\citenamefont {Liu}\ \emph {et~al.}(2019)\citenamefont {Liu},
  \citenamefont {Wang}, \citenamefont {Sato}, \citenamefont {Hohenadler},
  \citenamefont {Wang}, \citenamefont {Guo},\ and\ \citenamefont
  {Assaad}}]{Liu:2019aa}%
  \BibitemOpen
  \bibfield  {author} {\bibinfo {author} {\bibfnamefont {Y.}~\bibnamefont
  {Liu}}, \bibinfo {author} {\bibfnamefont {Z.}~\bibnamefont {Wang}}, \bibinfo
  {author} {\bibfnamefont {T.}~\bibnamefont {Sato}}, \bibinfo {author}
  {\bibfnamefont {M.}~\bibnamefont {Hohenadler}}, \bibinfo {author}
  {\bibfnamefont {C.}~\bibnamefont {Wang}}, \bibinfo {author} {\bibfnamefont
  {W.}~\bibnamefont {Guo}}, \ and\ \bibinfo {author} {\bibfnamefont {F.~F.}\
  \bibnamefont {Assaad}},\ }\href {\doibase 10.1038/s41467-019-10372-0}
  {\bibfield  {journal} {\bibinfo  {journal} {Nature Commun.}\ }\textbf
  {\bibinfo {volume} {10}},\ \bibinfo {pages} {2658} (\bibinfo {year}
  {2019})}\BibitemShut {NoStop}%
\bibitem [{\citenamefont {Wang}\ \emph {et~al.}(2021)\citenamefont {Wang},
  \citenamefont {Liu}, \citenamefont {Sato}, \citenamefont {Hohenadler},
  \citenamefont {Wang}, \citenamefont {Guo},\ and\ \citenamefont
  {Assaad}}]{Zwang_doping}%
  \BibitemOpen
  \bibfield  {author} {\bibinfo {author} {\bibfnamefont {Z.}~\bibnamefont
  {Wang}}, \bibinfo {author} {\bibfnamefont {Y.}~\bibnamefont {Liu}}, \bibinfo
  {author} {\bibfnamefont {T.}~\bibnamefont {Sato}}, \bibinfo {author}
  {\bibfnamefont {M.}~\bibnamefont {Hohenadler}}, \bibinfo {author}
  {\bibfnamefont {C.}~\bibnamefont {Wang}}, \bibinfo {author} {\bibfnamefont
  {W.}~\bibnamefont {Guo}}, \ and\ \bibinfo {author} {\bibfnamefont {F.~F.}\
  \bibnamefont {Assaad}},\ }\href {\doibase 10.1103/PhysRevLett.126.205701}
  {\bibfield  {journal} {\bibinfo  {journal} {Phys. Rev. Lett.}\ }\textbf
  {\bibinfo {volume} {126}},\ \bibinfo {pages} {205701} (\bibinfo {year}
  {2021})}\BibitemShut {NoStop}%
\bibitem [{\citenamefont {Kane}\ and\ \citenamefont
  {Mele}(2005)}]{KaneMele05b}%
  \BibitemOpen
  \bibfield  {author} {\bibinfo {author} {\bibfnamefont {C.~L.}\ \bibnamefont
  {Kane}}\ and\ \bibinfo {author} {\bibfnamefont {E.~J.}\ \bibnamefont
  {Mele}},\ }\href {\doibase 10.1103/PhysRevLett.95.146802} {\bibfield
  {journal} {\bibinfo  {journal} {Phys. Rev. Lett.}\ }\textbf {\bibinfo
  {volume} {95}},\ \bibinfo {pages} {146802} (\bibinfo {year}
  {2005})}\BibitemShut {NoStop}%
\bibitem [{\citenamefont {Hou}\ \emph {et~al.}(2023)\citenamefont {Hou},
  \citenamefont {Liu}, \citenamefont {Sato}, \citenamefont {Guo}, \citenamefont
  {Assaad},\ and\ \citenamefont {Wang}}]{EP_QSH}%
  \BibitemOpen
  \bibfield  {author} {\bibinfo {author} {\bibfnamefont {D.}~\bibnamefont
  {Hou}}, \bibinfo {author} {\bibfnamefont {Y.}~\bibnamefont {Liu}}, \bibinfo
  {author} {\bibfnamefont {T.}~\bibnamefont {Sato}}, \bibinfo {author}
  {\bibfnamefont {W.}~\bibnamefont {Guo}}, \bibinfo {author} {\bibfnamefont
  {F.~F.}\ \bibnamefont {Assaad}}, \ and\ \bibinfo {author} {\bibfnamefont
  {Z.}~\bibnamefont {Wang}},\ }\href {\doibase 10.1103/PhysRevB.107.155107}
  {\bibfield  {journal} {\bibinfo  {journal} {Phys. Rev. B}\ }\textbf {\bibinfo
  {volume} {107}},\ \bibinfo {pages} {155107} (\bibinfo {year}
  {2023})}\BibitemShut {NoStop}%
\bibitem [{Note1()}]{Note1}%
  \BibitemOpen
  \bibinfo {note} {From now on, we identify the K-IVC state in continuum
  description of MATBG as the dynamically generated QSH state in our model, and
  keep the name of K-IVC through the text.}\BibitemShut {Stop}%
\bibitem [{\citenamefont {Assaad}\ \emph {et~al.}(2022)\citenamefont {Assaad},
  \citenamefont {Bercx}, \citenamefont {Goth}, \citenamefont {G{\"o}tz},
  \citenamefont {Hofmann}, \citenamefont {Huffman}, \citenamefont {Liu},
  \citenamefont {Toldin}, \citenamefont {Portela},\ and\ \citenamefont
  {Schwab}}]{ALF_v2}%
  \BibitemOpen
  \bibfield  {author} {\bibinfo {author} {\bibfnamefont {F.~F.}\ \bibnamefont
  {Assaad}}, \bibinfo {author} {\bibfnamefont {M.}~\bibnamefont {Bercx}},
  \bibinfo {author} {\bibfnamefont {F.}~\bibnamefont {Goth}}, \bibinfo {author}
  {\bibfnamefont {A.}~\bibnamefont {G{\"o}tz}}, \bibinfo {author}
  {\bibfnamefont {J.~S.}\ \bibnamefont {Hofmann}}, \bibinfo {author}
  {\bibfnamefont {E.}~\bibnamefont {Huffman}}, \bibinfo {author} {\bibfnamefont
  {Z.}~\bibnamefont {Liu}}, \bibinfo {author} {\bibfnamefont {F.~P.}\
  \bibnamefont {Toldin}}, \bibinfo {author} {\bibfnamefont {J.~S.~E.}\
  \bibnamefont {Portela}}, \ and\ \bibinfo {author} {\bibfnamefont
  {J.}~\bibnamefont {Schwab}},\ }\href {\doibase 10.21468/SciPostPhysCodeb.1}
  {\bibfield  {journal} {\bibinfo  {journal} {SciPost Phys. Codebases}\ ,\
  \bibinfo {pages} {1}} (\bibinfo {year} {2022})}\BibitemShut {NoStop}%
\bibitem [{\citenamefont {Blankenbecler}\ \emph {et~al.}(1981)\citenamefont
  {Blankenbecler}, \citenamefont {Scalapino},\ and\ \citenamefont
  {Sugar}}]{Blankenbecler81}%
  \BibitemOpen
  \bibfield  {author} {\bibinfo {author} {\bibfnamefont {R.}~\bibnamefont
  {Blankenbecler}}, \bibinfo {author} {\bibfnamefont {D.~J.}\ \bibnamefont
  {Scalapino}}, \ and\ \bibinfo {author} {\bibfnamefont {R.~L.}\ \bibnamefont
  {Sugar}},\ }\href {\doibase 10.1103/PhysRevD.24.2278} {\bibfield  {journal}
  {\bibinfo  {journal} {Phys. Rev. D}\ }\textbf {\bibinfo {volume} {24}},\
  \bibinfo {pages} {2278} (\bibinfo {year} {1981})}\BibitemShut {NoStop}%
\bibitem [{\citenamefont {White}\ \emph {et~al.}(1989)\citenamefont {White},
  \citenamefont {Scalapino}, \citenamefont {Sugar}, \citenamefont {Loh},
  \citenamefont {Gubernatis},\ and\ \citenamefont {Scalettar}}]{White89}%
  \BibitemOpen
  \bibfield  {author} {\bibinfo {author} {\bibfnamefont {S.}~\bibnamefont
  {White}}, \bibinfo {author} {\bibfnamefont {D.}~\bibnamefont {Scalapino}},
  \bibinfo {author} {\bibfnamefont {R.}~\bibnamefont {Sugar}}, \bibinfo
  {author} {\bibfnamefont {E.}~\bibnamefont {Loh}}, \bibinfo {author}
  {\bibfnamefont {J.}~\bibnamefont {Gubernatis}}, \ and\ \bibinfo {author}
  {\bibfnamefont {R.}~\bibnamefont {Scalettar}},\ }\href {\doibase
  10.1103/PhysRevB.40.506} {\bibfield  {journal} {\bibinfo  {journal} {Phys.
  Rev. B}\ }\textbf {\bibinfo {volume} {40}},\ \bibinfo {pages} {506} (\bibinfo
  {year} {1989})}\BibitemShut {NoStop}%
\bibitem [{\citenamefont {Assaad}\ and\ \citenamefont
  {Evertz}(2008)}]{Assaad08_rev}%
  \BibitemOpen
  \bibfield  {author} {\bibinfo {author} {\bibfnamefont {F.}~\bibnamefont
  {Assaad}}\ and\ \bibinfo {author} {\bibfnamefont {H.}~\bibnamefont
  {Evertz}},\ }in\ \href {\doibase 10.1007/978-3-540-74686-7_10} {\emph
  {\bibinfo {booktitle} {Computational Many-Particle Physics}}},\ \bibinfo
  {series} {Lecture Notes in Physics}, Vol.\ \bibinfo {volume} {739},\ \bibinfo
  {editor} {edited by\ \bibinfo {editor} {\bibfnamefont {H.}~\bibnamefont
  {Fehske}}, \bibinfo {editor} {\bibfnamefont {R.}~\bibnamefont {Schneider}}, \
  and\ \bibinfo {editor} {\bibfnamefont {A.}~\bibnamefont {Wei{\ss}e}}}\
  (\bibinfo  {publisher} {Springer},\ \bibinfo {address} {Berlin Heidelberg},\
  \bibinfo {year} {2008})\ pp.\ \bibinfo {pages} {277--356}\BibitemShut
  {NoStop}%
\bibitem [{\citenamefont {Wu}\ and\ \citenamefont {Zhang}(2005)}]{Wu04}%
  \BibitemOpen
  \bibfield  {author} {\bibinfo {author} {\bibfnamefont {C.}~\bibnamefont
  {Wu}}\ and\ \bibinfo {author} {\bibfnamefont {S.-C.}\ \bibnamefont {Zhang}},\
  }\href {\doibase 10.1103/PhysRevB.71.155115} {\bibfield  {journal} {\bibinfo
  {journal} {Phys. Rev. B}\ }\textbf {\bibinfo {volume} {71}},\ \bibinfo
  {pages} {155115} (\bibinfo {year} {2005})}\BibitemShut {NoStop}%
\bibitem [{\citenamefont {Sugiyama}\ and\ \citenamefont
  {Koonin}(1986)}]{Sugiyama86}%
  \BibitemOpen
  \bibfield  {author} {\bibinfo {author} {\bibfnamefont {G.}~\bibnamefont
  {Sugiyama}}\ and\ \bibinfo {author} {\bibfnamefont {S.}~\bibnamefont
  {Koonin}},\ }\href {\doibase http://dx.doi.org/10.1016/0003-4916(86)90107-7}
  {\bibfield  {journal} {\bibinfo  {journal} {Annals of Physics}\ }\textbf
  {\bibinfo {volume} {168}},\ \bibinfo {pages} {1 } (\bibinfo {year}
  {1986})}\BibitemShut {NoStop}%
\bibitem [{\citenamefont {Sorella}\ \emph {et~al.}(1989)\citenamefont
  {Sorella}, \citenamefont {Baroni}, \citenamefont {Car},\ and\ \citenamefont
  {Parrinello}}]{Sorella89}%
  \BibitemOpen
  \bibfield  {author} {\bibinfo {author} {\bibfnamefont {S.}~\bibnamefont
  {Sorella}}, \bibinfo {author} {\bibfnamefont {S.}~\bibnamefont {Baroni}},
  \bibinfo {author} {\bibfnamefont {R.}~\bibnamefont {Car}}, \ and\ \bibinfo
  {author} {\bibfnamefont {M.}~\bibnamefont {Parrinello}},\ }\href
  {http://stacks.iop.org/0295-5075/8/i=7/a=014} {\bibfield  {journal} {\bibinfo
   {journal} {EPL (Europhysics Letters)}\ }\textbf {\bibinfo {volume} {8}},\
  \bibinfo {pages} {663} (\bibinfo {year} {1989})}\BibitemShut {NoStop}%
\bibitem [{Note2()}]{Note2}%
  \BibitemOpen
  \bibinfo {note} {For a finite size calculation in the canonical ensemble, one
  may need a very large system to observe phase separation due to the
  energetics of the domain boundary.}\BibitemShut {Stop}%
\bibitem [{\citenamefont {Fisher}\ \emph {et~al.}(1989)\citenamefont {Fisher},
  \citenamefont {Weichman}, \citenamefont {Grinstein},\ and\ \citenamefont
  {Fisher}}]{Fisher1989}%
  \BibitemOpen
  \bibfield  {author} {\bibinfo {author} {\bibfnamefont {M.~P.~A.}\
  \bibnamefont {Fisher}}, \bibinfo {author} {\bibfnamefont {P.~B.}\
  \bibnamefont {Weichman}}, \bibinfo {author} {\bibfnamefont {G.}~\bibnamefont
  {Grinstein}}, \ and\ \bibinfo {author} {\bibfnamefont {D.~S.}\ \bibnamefont
  {Fisher}},\ }\href {\doibase 10.1103/PhysRevB.40.546} {\bibfield  {journal}
  {\bibinfo  {journal} {Phys. Rev. B}\ }\textbf {\bibinfo {volume} {40}},\
  \bibinfo {pages} {546} (\bibinfo {year} {1989})}\BibitemShut {NoStop}%
\bibitem [{\citenamefont {Hasenbusch}(2019)}]{Hasenbusch2019}%
  \BibitemOpen
  \bibfield  {author} {\bibinfo {author} {\bibfnamefont {M.}~\bibnamefont
  {Hasenbusch}},\ }\href {\doibase 10.1103/PhysRevB.100.224517} {\bibfield
  {journal} {\bibinfo  {journal} {Phys. Rev. B}\ }\textbf {\bibinfo {volume}
  {100}},\ \bibinfo {pages} {224517} (\bibinfo {year} {2019})}\BibitemShut
  {NoStop}%
\bibitem [{\citenamefont {{Beach}}(2004)}]{Beach04a}%
  \BibitemOpen
  \bibfield  {author} {\bibinfo {author} {\bibfnamefont {K.~S.~D.}\
  \bibnamefont {{Beach}}},\ }\href@noop {} {\bibfield  {journal} {\bibinfo
  {journal} {eprint arXiv:cond-mat/0403055}\ } (\bibinfo {year} {2004})},\
  \Eprint {http://arxiv.org/abs/cond-mat/0403055} {cond-mat/0403055}
  \BibitemShut {NoStop}%
\bibitem [{\citenamefont {Bistritzer}\ and\ \citenamefont
  {MacDonald}(2011)}]{Bistritzer11}%
  \BibitemOpen
  \bibfield  {author} {\bibinfo {author} {\bibfnamefont {R.}~\bibnamefont
  {Bistritzer}}\ and\ \bibinfo {author} {\bibfnamefont {A.~H.}\ \bibnamefont
  {MacDonald}},\ }\href {\doibase 10.1073/pnas.1108174108} {\bibfield
  {journal} {\bibinfo  {journal} {Proceedings of the National Academy of
  Sciences}\ }\textbf {\bibinfo {volume} {108}},\ \bibinfo {pages} {12233}
  (\bibinfo {year} {2011})},\ \Eprint
  {http://arxiv.org/abs/https://www.pnas.org/doi/pdf/10.1073/pnas.1108174108}
  {https://www.pnas.org/doi/pdf/10.1073/pnas.1108174108} \BibitemShut {NoStop}%
\bibitem [{\citenamefont {Zhang}\ \emph {et~al.}(2021)\citenamefont {Zhang},
  \citenamefont {Pan}, \citenamefont {Zhang}, \citenamefont {Kang},\ and\
  \citenamefont {Meng}}]{ZhangX21}%
  \BibitemOpen
  \bibfield  {author} {\bibinfo {author} {\bibfnamefont {X.}~\bibnamefont
  {Zhang}}, \bibinfo {author} {\bibfnamefont {G.}~\bibnamefont {Pan}}, \bibinfo
  {author} {\bibfnamefont {Y.}~\bibnamefont {Zhang}}, \bibinfo {author}
  {\bibfnamefont {J.}~\bibnamefont {Kang}}, \ and\ \bibinfo {author}
  {\bibfnamefont {Z.~Y.}\ \bibnamefont {Meng}},\ }\href {\doibase
  10.1088/0256-307X/38/7/077305} {\bibfield  {journal} {\bibinfo  {journal}
  {Chinese Physics Letters}\ }\textbf {\bibinfo {volume} {38}},\ \bibinfo
  {pages} {077305} (\bibinfo {year} {2021})}\BibitemShut {NoStop}%
\bibitem [{\citenamefont {Hofmann}\ \emph {et~al.}(2022)\citenamefont
  {Hofmann}, \citenamefont {Khalaf}, \citenamefont {Vishwanath}, \citenamefont
  {Berg},\ and\ \citenamefont {Lee}}]{Hofmann22}%
  \BibitemOpen
  \bibfield  {author} {\bibinfo {author} {\bibfnamefont {J.~S.}\ \bibnamefont
  {Hofmann}}, \bibinfo {author} {\bibfnamefont {E.}~\bibnamefont {Khalaf}},
  \bibinfo {author} {\bibfnamefont {A.}~\bibnamefont {Vishwanath}}, \bibinfo
  {author} {\bibfnamefont {E.}~\bibnamefont {Berg}}, \ and\ \bibinfo {author}
  {\bibfnamefont {J.~Y.}\ \bibnamefont {Lee}},\ }\href {\doibase
  10.1103/PhysRevX.12.011061} {\bibfield  {journal} {\bibinfo  {journal} {Phys.
  Rev. X}\ }\textbf {\bibinfo {volume} {12}},\ \bibinfo {pages} {011061}
  (\bibinfo {year} {2022})}\BibitemShut {NoStop}%
\end{thebibliography}%

\clearpage

\section{Supplemental Material}

\subsection{Data for $\Delta=0.75$}

\begin{figure}[htbp]
\centering
\includegraphics[width=0.41\textwidth]{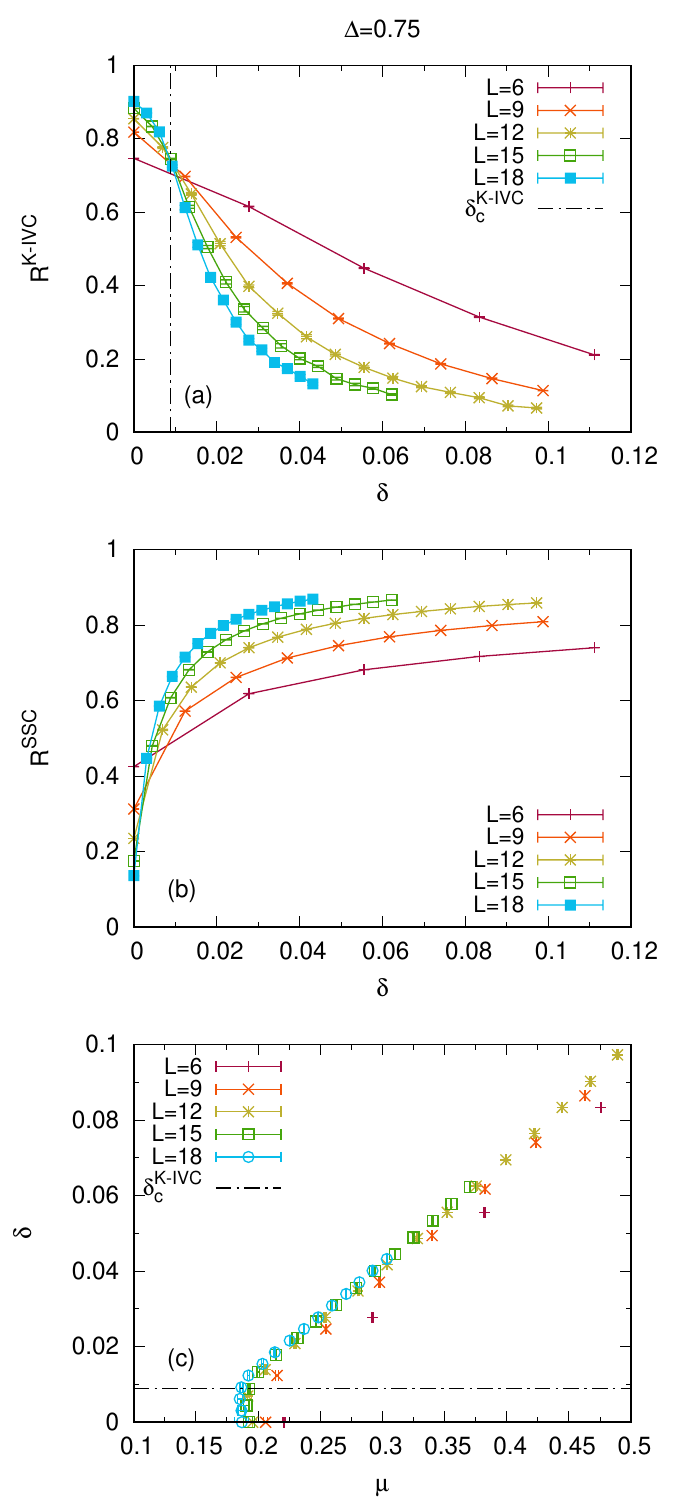}
\caption{\label{fig:Ratioeq_0.75}  
Data for $\Delta=0.75$. 
Correlation ratios for (a) $R^{\rm K-IVC}$ and (b)  $R^{\rm SSC}$ as a function of doping factor $\delta$. (c)  $\delta$ as a function of chemical potential $\mu$. The vertical and horizontal dashed lines are a guide to the eye, fitted from the crossing point analysis of the K-IVC order parameter.   Instead of choosing the largest two sizes, we fit the crossing points between lines of $L$ and $L+3$ from $L=6$ to $L=15$ and extrapolate to the thermodynamic limit. 
}
\end{figure}

Here  we   present  data  for  $\Delta=0.75$.
Upon  inspection of   Fig.~\ref{fig:Ratioeq_0.75}(a)-(c)   the  data  bears very similarities  to  the  case of $\Delta=0.5$ 
presented in the main  text.    In particular,  we  observe a  first-order  phase  transition 
between the K-IVC  and  SSC.  
However  in  comparison to  the  case  $\Delta = 0.5$  we   see  that the doping range where  phase 
separation occurs   $  0  < \delta    <  \delta^{\text{K-IVC}}_c$    is  smaller.

\subsection{Single-particle spectrum} 

\begin{figure}[htbp]
  \centering
  \includegraphics[width=0.5\textwidth]{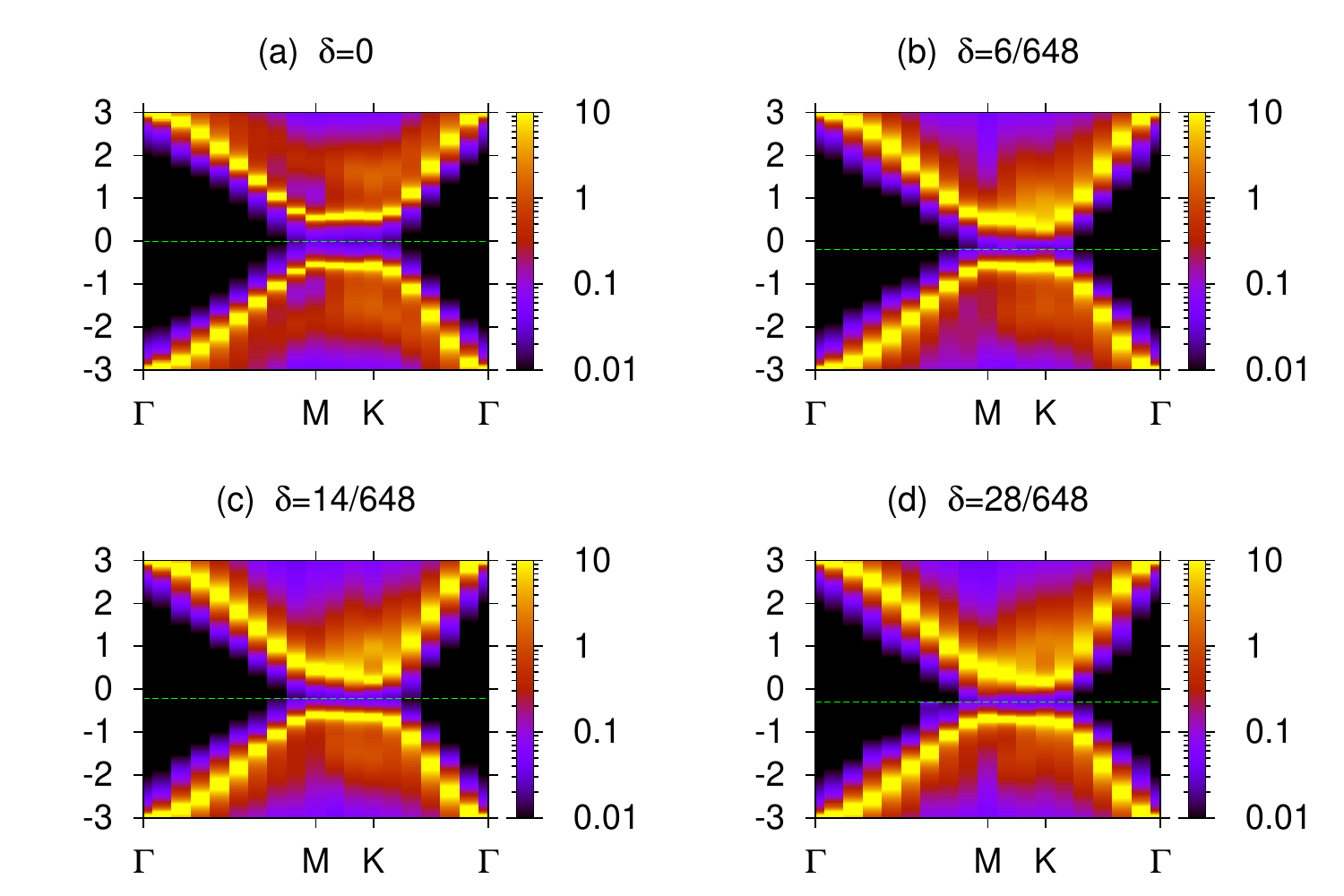}
  \caption{\label{fig:spL18-SM}
Single-particle spectrum for several dopings at $\Delta=0.75$.  
(a) $\delta=0$ (inside K-IVC phase); (b) $\delta=\frac{6}{648}$ (around K-IVC-SSC transition); (c) $\delta=\frac{14}{648}$ (inside SSC phase); and (d) $\delta=\frac{28}{648}$ (inside SSC phase). 
The green dotted line corresponds to the  chemical potential $ \mu $ evaluated from
Eq.~(\ref{Eq:cp-pqmc}) of the main text.}
\end{figure}

In this section, we show the   single-particle spectrum at finite chemical potential.   At  zero  temperature  we  can sharply  
 distinguish between the  particle  addition  and  removal  spectra:  
\begin{equation}
\begin{aligned} 
 & A(\omega)=A_{+}(\omega) + A_{-}(-\omega) \\ 
 &  \langle  c_{\bm{k}, \sigma} (\tau)  c^{\dagger}_{ \bm{k}, \sigma }(0) \rangle  =  \int d\omega  e^{ -\tau \omega }  A_{+}(\omega )      \\
 &  \langle  c^{\dagger}_{\bm{k}, \sigma} (\tau)  c_{ \bm{k}, \sigma }(0) \rangle  =  \int d\omega  e^{ -\tau \omega }  A_{-}(\omega ).  
\end{aligned}     
\end{equation}
We  obtain  the spectral  function    by  analytic continuation  in particle and hole channels:  
\begin{equation}
\begin{aligned}\label{Eq:Spectrum}
A(\bm{k}, \omega) & = \frac{1}{Z}  \sum_n
  ( | \langle  n | c_{\bm{k}} | 0 \rangle |^2  \delta(E_n - E_0 - \omega)  )  \\
& +  \frac{1}{Z}  \sum_m ( | \langle  m | c^{\dagger}_{\bm{k}} | 0 \rangle |^2  \delta(E_m - E_0 + \omega)  ).
\end{aligned}
\end{equation}
Here, $|0 \rangle$ in
Eq.~(\ref{Eq:Spectrum}) is the ground state at finite doping and $\langle n |$
is an eigenstate of the Hamiltonian with energy $E_n$ and an additional particle
(hole) relative to the ground state.    We  have  used  the  stochastic  Maxent   \cite{Beach04a}    implementation 
of  the  ALF \cite{ALF_v2}   library. 

In
Fig.~\ref{fig:spL18-SM}, we plot the spectral functions for   the $L=18$  lattice   and
 dopings,  $\delta=0, \frac{6}{648}, \frac{14}{648}$, and $\frac{28}{648}$.  
As  apparent,  we always  observe a   single-particle   gap   in the spectra and an approximate particle-hole symmetry around the Fermi level.    At  finite  doping,   this  stems  
from the  superconducting nature of  the  ground state  and  the energy  cost of  breaking a Cooper  pair. 
Importantly, since fermion excitations  are  gapped, one  can integrate  them  out,   to   derive  the 
purely  bosonic  field  theory    discussed  in   Ref.~\cite{Khalaf21}.

\subsection{Preformed pairs }

Here,  we would like to clarify  two points:  
i)  pairs of merons are the lowest energy excitations  and  
ii) the   gap of preformed pairs  increases as a function of anisotropy.

 We extrapolate the  excitation energy of  the pairing gap $\Delta_{\text{SC}}$   as  obtained from
the SSC imaginary-time correlations, 
\begin{equation}
\begin{aligned}
  &	\frac{1}{L^2} \sum_{\bm{r, r'}}
	\left[ \langle  \hat{ \eta }^{+}_{\boldsymbol{r},\boldsymbol{ \widetilde{\delta} }_a }(\tau)  \hat{ \eta  }^{-}_{\boldsymbol{r'},\boldsymbol{ \widetilde{ \delta} }_b }(0)  \rangle +
	 \langle  \hat{ \eta }^{- }_{\boldsymbol{r},\boldsymbol{ \widetilde{\delta} }_a }(\tau)  \hat{ \eta  }^{+}_{\boldsymbol{r'},\boldsymbol{ \widetilde{\delta} }_b }(0)    \rangle\right]\\
& \propto  e^{ -\Delta_{ \text{SC} } \tau } ,
\end{aligned}
\end{equation}
as  well as   the fermionic single-particle gap from the Green's function:
\begin{equation}
\begin{aligned}
  \sum_{\sigma} \langle  c_{\bm{k},\sigma } (\tau)  c^{\dagger}_{ \bm{k}, \sigma }(0) \rangle  \propto  e^{ -\Delta_{ \text{sp} } \tau }.
\end{aligned}
\end{equation}
We  note  that  the minimal gap   is  at  the momentum $M$ point in the  Brillouin  zone (see  Fig.~\ref{fig:spL18-SM}).

\begin{figure}[htbp]
\centering
\includegraphics[width=0.41\textwidth]{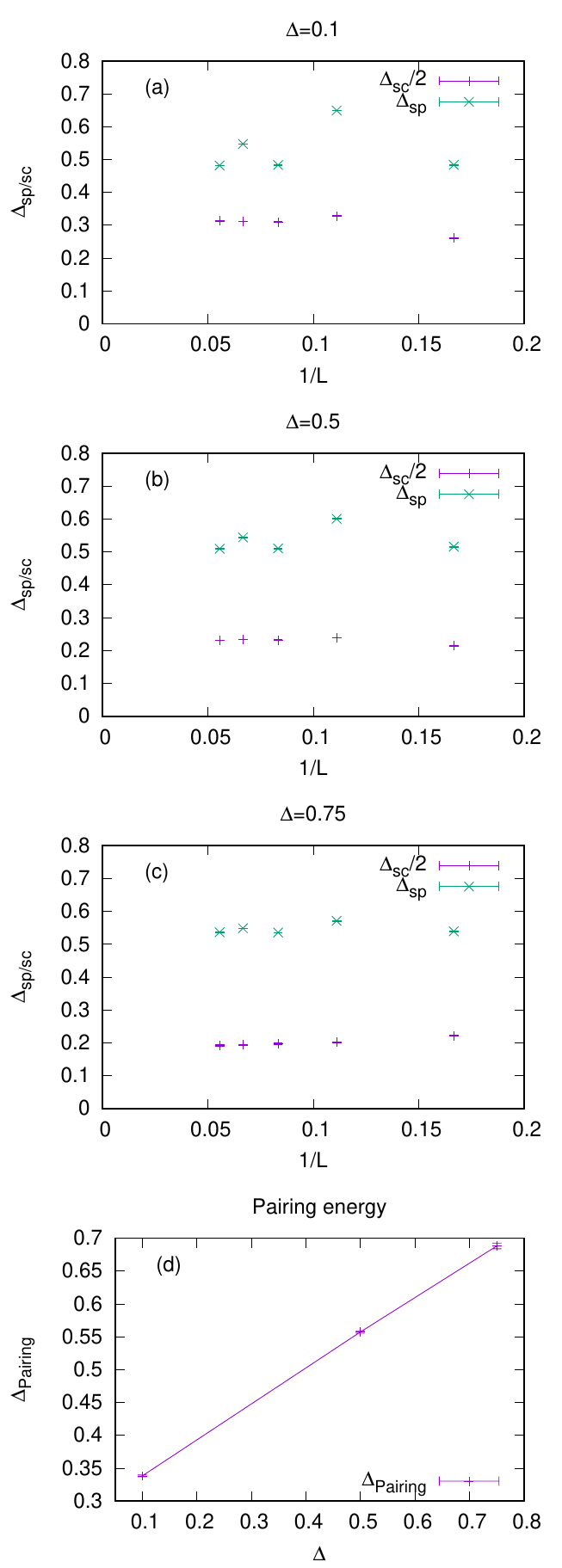}
\caption{\label{fig:Gap_Meron}
Pairing gap and single-particle gap as a function of $1/L$ in the K-IVC state for three cases: 
(a) $\Delta=0.1$, $\lambda=0.043$; (b) $\Delta=0.5$, $\lambda=0.035$; and
(c) $\Delta=0.75$, $\lambda=0.03$.  The gap of preformed pairs $ \Delta_{\text{Pairing}} \equiv 2 \Delta_{\text{sp}} - \Delta_{\text{SC}} $ 
 for $L=18$ is shown in Fig.~(d).}
\end{figure}

The estimated finite-size pairing gaps and single-particle gaps for sizes $L=6,9,...,18$  are shown
in Fig.~\ref{fig:Gap_Meron}.  We  consider   values of $\lambda$  that  lie deep   within  the K-IVC phase:
$\lambda=0.03$ for $\Delta=0.75$, $\lambda=0.035$ for $\Delta=0.5$ , and $\lambda=0.043$ for $\Delta=0.1$.
For odd $L$ ($9$ and $15$), we select the nearest
 momentum  around the $M$ point to calculate $\Delta_{\rm sp}$.  This  explains  the  even-odd oscillations 
 in  the finite-size estimators (see  Fig.~\ref{fig:Gap_Meron}). 

Overall, the  single-particle gap $\Delta_{\rm sp}$ is  larger than half of the pairing gap $\Delta_{\rm SC}/2$ for all three values of $\Delta$, 
 thus  indicating  that  the  lowest  charge excitation  is  a  pair.    
The  pairing  energy   is  defined  as   $ \Delta_{ \text{Pairing} }  = 2\Delta_{\text{sp}} - \Delta_{\text{SC}} $.  
Let us now  consider  only  the even lattice  sizes  for  which  the   single-particle gap  shows  
little  size effects.  In this  case,  we  observe  that the pairing  energy  decreases  as  the    easy-plane 
anisotropy   grows. 
This is consistent with
our understanding based on the meron picture:  although  the  norm of  the  K-IVC U(1) order parameter (corresponding 
to  the   single-particle  gap)    has no
significant dependence on the  anisotropy,  pairs of merons require less excitation energy as  the  anisotropy  grows.  This  stems  
from our  understanding  that as  the anisotropy  grows  the  meron  core  becomes more energetically   expensive.

\subsection{Spin current texture} 

The fact that  
 skyrmions in $S^2$ space carry charge $2e$ can be understood locally:  
 electron  densities are associated with the curvature of  the  local spin-orbit fluctuations.  
 Correspondingly,  spin current textures will also be observed  when the  electron (hole)  
 distribution is not uniform.  
In this section,   
we  will demonstrate numerically the one-to-one relation 
between these two quantities.

\begin{figure}[htbp]
\centering
\includegraphics[width=0.45\textwidth]{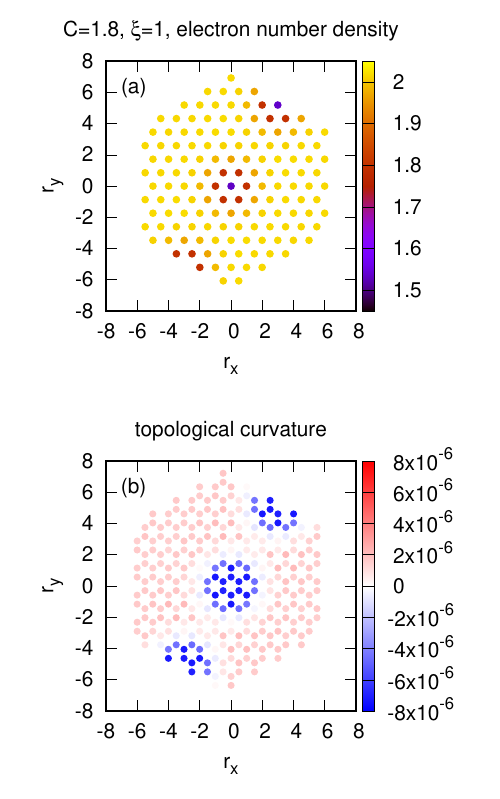}
\caption{\label{fig:pontryagin}
(a)   Electron particle number distribution in each unit cell, and (b) the curvature in $S^2$ space of spin current  in the presence of a finite pinning potential based on Eq.~(\ref{Eq:Pinning}).  The simulation is 
  performed on a $12 \times 12$ lattice  with  $\delta  =   2/288$, $\Delta=0.75$, and $\lambda=0.03$.  
}
\end{figure}

We trap a pair of holes at two separate regions of the honeycomb lattice, by adding a site-dependent pinning 
potential:  
\begin{equation}
\begin{aligned}\label{Eq:Pinning}
H_{pin} =& C \sum_{\boldsymbol r} \sum_{\tilde{\boldsymbol \delta}}
\exp(-|\boldsymbol r - \bm r_{c_1} +\tilde{\boldsymbol \delta}|/\xi){\hat c}^\dagger_{\boldsymbol r 
 +\tilde{\boldsymbol \delta}}{\hat c}^{}_{\boldsymbol r+\tilde{\boldsymbol \delta}} \\
 + & C \sum_{\boldsymbol r} \sum_{\tilde{\boldsymbol \delta}} \exp(-|\boldsymbol r-\boldsymbol r_{c_2}+\tilde{\boldsymbol \delta}|/\xi){\hat c}^\dagger_{\boldsymbol r+\tilde{\boldsymbol \delta}}{\hat c}^{}_{\boldsymbol r+\tilde{\boldsymbol \delta}}    
\end{aligned}
\end{equation}
such that the electron density relative to half-filling is reduced around the center  
of two  potential wells $ \bm{r}_{c1} $ and $\bm{r}_{c2} $, as shown by Fig.~\ref{fig:pontryagin}(a).

\begin{figure}[htbp]
\centering
\includegraphics[width=0.36\textwidth]{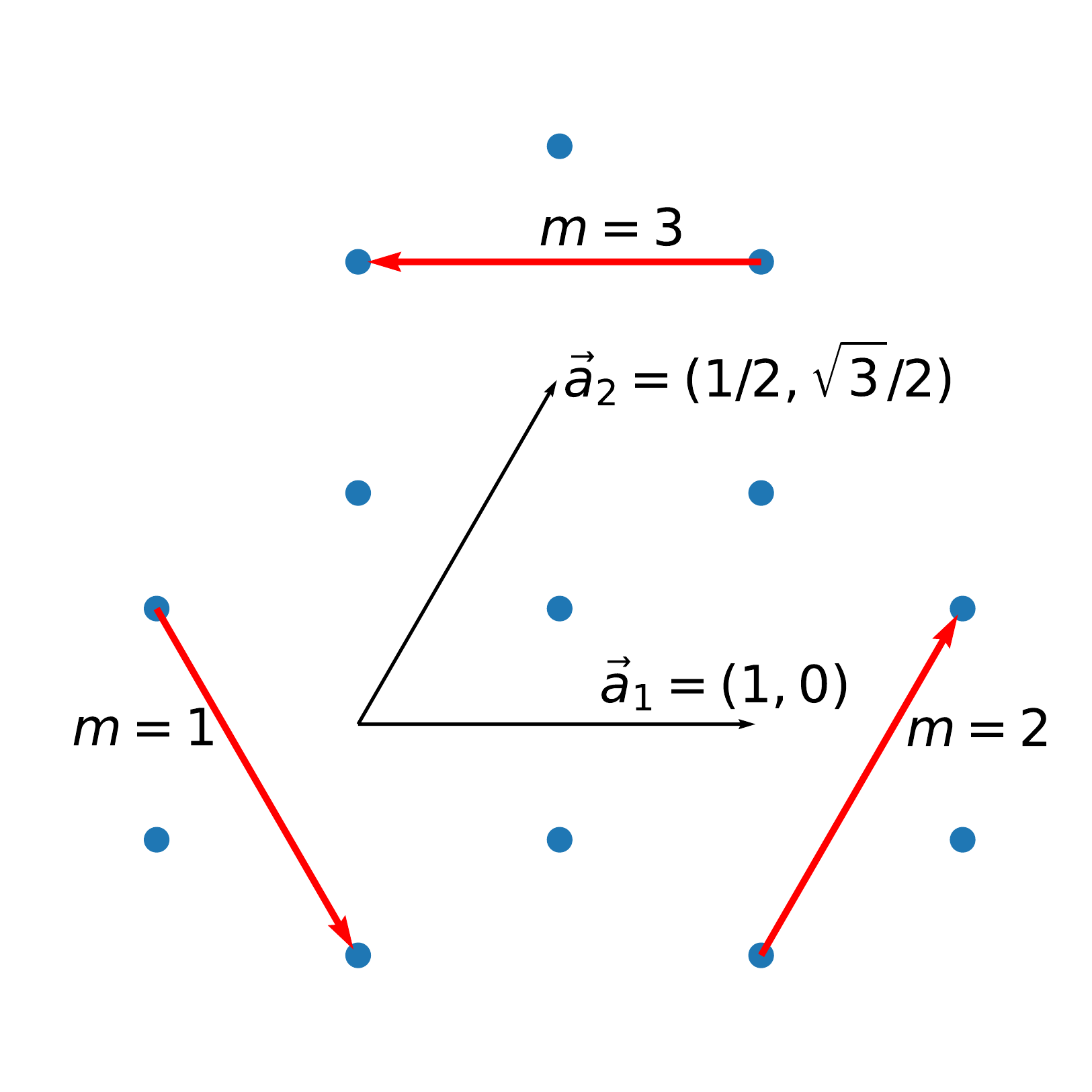}
\caption{\label{fig:honeycomb}
 Definition of the three next-nearest-neighbor bonds $m=1,2,3$ used for defining the curvature operator in Eq.~(\ref{def:curv}).  
}
\end{figure}

Correspondingly,  the curvature of the spin-orbit order parameter is  given by 
\begin{equation}
\begin{aligned}
 C_{ \bm{r} } 
 &= {\frac{1}{4\pi}}  \hat{\bm J}_{ \bm{r} } \cdot 
  ( \partial_x  \hat{\bm J}_{\bm{r}} \times  \partial_y \hat{\bm J}_{\bm{r}})    \\ 
 & =  \frac{1}{ 4\pi \mathcal{A} } \det 
  \begin{pmatrix} 
   \hat{J}^X_{ \bm{r}, m=1 } &  \hat{J}^X_{ \bm{r} + \bm{a}_1, m=2 } &   \hat{J}^X_{ \bm{r} + \bm{a}_2, m=3 }  \\  
   \hat{J}^Y_{ \bm{r}, m=1 } &  \hat{J}^Y_{ \bm{r} + \bm{a}_1, m=2 } &   \hat{J}^Y_{ \bm{r} + \bm{a}_2, m=3 }  \\  
   \hat{J}^Z_{ \bm{r}, m=1 } &  \hat{J}^Z_{ \bm{r} + \bm{a}_1, m=2 } &   \hat{J}^Z_{ \bm{r} + \bm{a}_2, m=3 }  
   \end{pmatrix}
\end{aligned} \label{def:curv}
\end{equation} 
where $\bm r$ denotes the position of unit cell and  
$m$ labels the six next nearest neighbor bonds. We  use $m=1,2,3$ respectively on hexagons at site $\bm r$, $\bm r + \bm a_1$ and $\bm r + \bm a_2$, as shown in Fig.~\ref{fig:honeycomb}. Here $\mathcal{A}$ is the area of a hexagon.   
This way of defining curvature  is to avoid explicit density operators via the commutator of spin current operators.
From Fig.~\ref{fig:pontryagin}(b), it is clear that the curvature distribution in real space follows  
exactly the spatial pattern of the electron density.

\subsection{Scaling analysis of the coupling between two order parameters }

The  aim of  this  section  is  to show  that at the  3D XY  critical point  coupling  to  charge  fluctuations  is  irrelevant. 
To this end, we write down the $\varphi^4$ theory  describing  the K-IVC phase transition  within the  superconducting state:
\begin{equation}
\begin{aligned}\label{Eq:phi4Theory}
 \mathcal{S} =&  \mathcal{S}_{\text{K-IVC}}  +  \mathcal{S}_{\text{SSC}}  + \mathcal{S}_{\text{coupling}},  \\
\end{aligned}
\end{equation}  
where $\mathcal{S}_{\text{K-IVC}}$ is the bare action for K-IVC field,  and $\mathcal{S}_{\text{SSC}}$ is the one for SSC order.  
Here we denote the two U(1) order parameters for 
K-IVC and SSC as ${\bm n}$ and ${ \phi}$, respectively. In the SSC phase, we can decompose the field  as 
$\phi(\bm x)=|\phi({\bm x})|\exp(i \theta({\bm x}))$ with $|\phi(\bm x)|=\phi_0+\phi_\parallel({\bm x})$. $\theta({\bm x})$ represents the Goldstone mode of $\phi$.

$\mathcal{S}_{\text{coupling}}$ describes  the charge- or spin-neutral coupling terms between two fields.  
 The leading term reads  
\begin{equation}
  g \int d^D r  | \bm n |^2  \partial_\mu \theta({\bm x})  \partial_\mu \theta({\bm x}),
\end{equation}
where $D=d+1$ and $d=2$. 
Note that terms like $\int d^D r  |\bm n|^2   |\phi|^2 $ will not  be present at this critical point,   
since $\phi$ has a finite mass along the longitudinal direction in  the  SSC phase.

Let $\Delta_n$ be the scaling dimension of  the field ${\bm n}$ and $\Delta_\theta$ be the scaling dimension of the  Goldstone boson ${ \theta}$.
According to standard scaling theory,  the scaling dimension of $\lambda$ should be $\Delta_\lambda=D-2\Delta_n-2(\Delta_\theta+1)$.

Assume that in our system, the Wilson-Fisher fixed point of the field ${\bm n}$ belongs  to  the  $2+1D$ universality class ( 
$\Delta_n=\frac{D-2+\eta}{2}=0.51905$, with $\eta=0.03810$).   
On the other hand,  the Goldstone mode $\theta$ of a superconducting state in $2+1$$D$  
is associated with a power-law decay correlation function along the transverse direction: 
\begin{equation}
  \langle e^{ i\theta({\bm r})} e^{i \theta(0)} \rangle  \propto \int \frac{d^3 k}{ (2\pi)^3 } \frac{ i \bm{k} \cdot \bm{r} }{ k^2 }   
   \propto  \frac{1}{|\bm{r}|}.
\end{equation}
This is based on the saddle point expansion of $\varphi^4$ theory along  the  massless direction.  
Therefore, $ \Delta_{\theta} = 0.5$, and as a consequence, 
$\Delta_\lambda=D-2\Delta_n-2\Delta_\theta -2 < 0$.    This implies that the coupling between two U(1) order parameters is 
irrelevant.    This is consistent with our numerical evidence of the $2+1$$D$  O(2) universality class   
at the K-IVC critical point.

\subsection{Relation to   magic-angle  twisted  bilayer  graphene (MATBG) } 

Our  model  is closely analogous to those   described  in Refs.~\cite{Bultinck20, Khalaf21}    that aim at  accounting  for    superconductivity resulting from the  condensation of skyrmions in   magic-angle  
twisted  bilayer  graphene  (MATBG). 
The  starting point is the    Bistritzer-MacDonald continuum  model \cite{Bistritzer11}  in the chiral  limit,  
in  which    interlayer  hopping   between  the   same  sub-lattice   is  set  to  zero.     In this  case,  the  low-lying   bands are  flat  
 and   can  be  labeled  by  sub-lattice  polarization, which is complete  in  this limit,  as  well as  a   valley  and  spin  index.   In this  basis,  the bands  
 carry  non-trivial topology   characterized by a Chern number.   
Let     $\hat{c}^{\dagger}_{\tau,\sigma, s }(\ve{k}) $    create an electron  with momentum $\ve{k} +  \ve{K}_{\tau}$,   in valley     $\tau$,  sub-lattice  polarization   $\sigma$,  and  physical   spin 
  $s$.    The  Chern  number of the bands  is  given  by  $  \sigma_z  \tau_z $.       In this  chiral  limit,      the   form  factor  of  the  
  density  fluctuations  depends  solely   on $ \sigma_z  \tau_z $,    such  that  the  complete  model  possesses  a   U(4) $ \times  $ U(4)    symmetry  that 
  rotates  the  bands  in a  given  Chern  sector.  This    model  captures  the  dominant  energy  scales.         Perturbations  beyond  the  chiral  limit  break  this symmetry.   
  In  fact,  the   phases   and   phase  transitions  discussed   in \cite{Bultinck20, Khalaf21}     are  captured  by  the  
  effective  model:
  \begin{eqnarray}
  \label{MATBG.eq}	
     \hat{H}  & & =   \sum_{\ve{k}}    \hat{c}^{\dagger}(\ve{k})  \left(  k_x  \sigma^{x}  +  k_y  \sigma^{y} \tau^{z} \right)   \hat{c}^{\phantom\dagger}(\ve{k})   \nonumber  \\    
     & &   - \lambda   \int_V  d^2{\ve{x}}  \hat{P}  \left(     \left[ \hat{c}^{\dagger}(\ve{x}) \tau^{x} \sigma^{y}  \hat{c}^{\phantom\dagger}(\ve{x})   \right]^2  
       \right.   \\
      & &  \left. +  \left[ \hat{c}^{\dagger}(\ve{x}) \tau^{y} \sigma^{y}  \hat{c}^{\phantom\dagger}(\ve{x})   \right]^2   + 
	     \Delta  \left[ \hat{c}^{\dagger}(\ve{x})  \sigma^{z}  \hat{c}^{\phantom\dagger}(\ve{x})   \right]^2 \right)   \hat{P}. \nonumber   
  \end{eqnarray} 
    Here,   $\hat{P} $    reflects the  projection onto the low energy  Hilbert  space.    To be more precise,  let 
   $  | \ve{k},  \tau,  \sigma, s \rangle  =    \hat{c}^{\dagger}_{\tau,\sigma, s }(\ve{k})  | 0 \rangle $  such  that  the  resolution 
   of  unity in the single-particle  Hilbert  space  reads:  $\hat{1} =  \sum_{\ve{k},  \tau,  \sigma, s } | \ve{k},  \tau,  \sigma, s \rangle  \langle   \ve{k},  \tau,  \sigma, s | +  \hat{P}_H $,  where  $\hat{P}_H$  denotes  the projection on the high  energy states. Then,  
$   \hat{c}^{\dagger}_{\tau,\sigma, s }(\ve{x})    =   \sum_{\ve{k}, \tau,\sigma,s}  \langle   \ve{k},  \tau,  \sigma, s |  
 \ve{x},  \tau,  \sigma, s  \rangle  \hat{c}^{\dagger}_{\tau,\sigma, s }(\ve{k})  $. 
As  mentioned  above, the single-particle wave functions of flat bands $| \ve{k},  \tau,  \sigma, s \rangle $ 
 have a characteristic Chern number of $\tau_z \sigma_z$.  
 Crucially, $   \hat{c}^{\dagger}_{\tau,\sigma, s }(\ve{x})  $  does not satisfy 
 the  fermion canonical  commutation rules since the transformation is not  unitary. 
  $ \ve{M}^{\text{VH}} = (  \tau^{x} \sigma^{y},  \tau^{y} \sigma^{y}, \sigma^{z} ) $  correspond  to  mutually   anti-commuting  
  mass  terms   that    account for the   so-called   Kramers intervalley-coherent (K-IVC)   and    valley-Hall (VH)    insulators.

      At   $\Delta = 1$   the  model  has  an  SU$_V$(2)   \textit{valley}   
  symmetry   with  generators    $  \frac{i}{2} \left[ M^{\text{VH}}_i, M^{\text{VH}}_j \right] $,   corresponding   to $ ( \tau^x \sigma^x, \tau^{y} \sigma^{x},\tau^{z})  $,       as  well  as   an SU$_S$(2)  spin symmetry and  a U$_C$(1) charge  symmetry. 
     For    $\Delta \ne 1 $,   the  SU$_V$(2)  
  symmetry  is  reduced  to  a U$_V$(1)  $\times$  Z$_2$  and for   $\Delta   < 1 $  the  K-IVC   state  is   favored.    Since  the  bands   have  a non-trivial  Chern index,  skyrmion  excitations  of  the    three-component   K-IVC   and  VH    order  parameters   carry    charge  $2e$.  
   We  note  that, since  the  
  SU$_V$(2)   is  reduced  to U$_V$(1)  $\times$  Z$_2$,  skyrmions  have to be seen  in terms of a  pair of   merons,  each carrying charge  $e$. 
  Hence, all in all,  the model   has  U$_V$(1)  $\times$  
  Z$_2$  $\times$  U$_C$(1)   $\times$  SU$_s$(2)  symmetry.  
    The  doping-induced  transition between the K-IVC  and  SSC  put  forward in  \cite{Khalaf21}  and  based on the proliferation of     skyrmions, 
       does  not involve  the  spin degrees   of  freedom.   Spinless  versions of  the model  have  been  put forward to  capture  the  relevant physics  \cite{Khalaf21, Ippoliti22}.

 The model  of  Eq.~\ref{MATBG.eq} does  not   support  lattice  regularization  since  it   will   break  the   valley   (or  chiral)   symmetry.   Furthermore,  
 the  topology of the  bands  does  not allow  for  a local Wannier   basis.  As a  result,  simulations  of  MATBG  are  carried  out in the  
 continuum  \cite{ZhangX21, Hofmann22, Ippoliti22}. 
 Our  model  provides   a  possibility  to   avoid  this   by    encoding   the    
 SU$_V$(2)     symmetry    as     SU$_S$(2).      A  continuum limit  of our  model  reads: 
   \begin{eqnarray}
  \label{QSHZ.eq}	
     \hat{H}  =  & &   \sum_{\ve{k}}    \hat{c}^{\dagger}(\ve{k})  \left(  k_x  \sigma^{x}  +   k_y  \sigma^{y}  \tau^{z} \right)   \hat{c}^{\phantom\dagger}(\ve{k})        \\  
     & &   - \lambda   \int_V  d^2{\ve{x}}   \left(     \left[ \hat{c}^{\dagger}(\ve{x})  s^{x} \tau^{z} \sigma^{z}  \hat{c}^{\phantom\dagger}(\ve{x})   \right]^2  
      \right.  \nonumber  \\
    & & 	\left. +  \left[ \hat{c}^{\dagger}(\ve{x}) s^{y} \tau^{z} \sigma^{z}  \hat{c}^{\phantom\dagger}(\ve{x})   \right]^2   + 
	     \Delta  \left[ \hat{c}^{\dagger}(\ve{x})  s^{z}   \tau^{z} \sigma^{z} \hat{c}^{\phantom\dagger}(\ve{x})   \right]^2  \right)  \nonumber, 
  \end{eqnarray}
  where   $ \ve{M}^{\text{QSH}} =  (s^x, s^y, s^z)   \tau^{z} \sigma^{z} $    correspond to the  three   QSH  mass  terms.      The parallel  now  becomes 
  apparent:  the   $ \ve{M}^{\text{QSH}}_z$  mass corresponds  to the  
  Valley  Hall  insulator,   and  the  first  two  components  to  the K-IVC  insulator.    Topologically,  both models  are  equivalent   since      the  skyrmion of  
  the     three-component  QSH  order  parameter  carries charge  $2e$.    The  symmetry of  the Hamiltonian   is  given  by  U$_s$(1)  $\times$  
  Z$_2$  $\times$  U$_C$(1)   $\times$  SU$_V$(2).     Here  the SU$_V$(2)   symmetry  is  again spanned  by  the generators: $ ( \tau^x \sigma^x, \tau^{y} \sigma^{x},\tau^{z})  $.

\end{document}